\documentclass[12p]{article}
\usepackage[margin=1in]{geometry}
\usepackage{xurl}
\usepackage[utf8]{inputenc}
\usepackage[round]{natbib}
\usepackage[pdftex]{graphicx}
\usepackage[export]{adjustbox}
\usepackage{xcolor}
\usepackage{textcomp}
\usepackage{gensymb}

\usepackage{verbatim}
\usepackage[section]{placeins}
\usepackage{nameref}
\usepackage{lineno}
\usepackage{setspace}
\usepackage{sectsty}
\usepackage{titling}
\usepackage{hyperref}
\hypersetup{
  colorlinks   = true,    % Colours links instead of boxes
  urlcolor     = blue,    % Colour for external hyperlinks
  linkcolor    = blue,    % Colour of internal links
  citecolor    = blue      % Colour of citations
}

\usepackage{etoolbox}

\makeatletter

\pretocmd{\NAT@citex}{%
  \let\NAT@hyper@\NAT@hyper@citex
  \def\NAT@postnote{#2}%
  \setcounter{NAT@total@cites}{0}%
  \setcounter{NAT@count@cites}{0}%
  \forcsvlist{\stepcounter{NAT@total@cites}\@gobble}{#3}}{}{}
\newcounter{NAT@total@cites}
\newcounter{NAT@count@cites}
\def\NAT@postnote{}

\def\NAT@hyper@citex#1{%
  \stepcounter{NAT@count@cites}%
  \hyper@natlinkstart{\@citeb\@extra@b@citeb}#1%
  \ifnumequal{\value{NAT@count@cites}}{\value{NAT@total@cites}}
    {\ifNAT@swa\else\if*\NAT@postnote*\else%
     \NAT@cmt\NAT@postnote\global\def\NAT@postnote{}\fi\fi}{}%
  \ifNAT@swa\else\if\relax\NAT@date\relax
  \else\NAT@@close\global\let\NAT@nm\@empty\fi\fi% avoid compact citations
  \hyper@natlinkend}
\renewcommand\hyper@natlinkbreak[2]{#1}

\patchcmd{\NAT@citex}
  {\ifNAT@swa\else\if*#2*\else\NAT@cmt#2\fi
   \if\relax\NAT@date\relax\else\NAT@@close\fi\fi}{}{}{}
\patchcmd{\NAT@citex}
  {\if\relax\NAT@date\relax\NAT@def@citea\else\NAT@def@citea@close\fi}
  {\if\relax\NAT@date\relax\NAT@def@citea\else\NAT@def@citea@space\fi}{}{}

\makeatother

\subsectionfont{\normalfont\itshape}

\title{Equivalent near-field corner frequency analysis of 3D dynamic rupture simulations reveals source complexity}
\author{Nico Schliwa$^1$ and Alice-Agnes Gabriel$^{2}$}
\date{}

\begin{document}

\maketitle

(1) Department of Earth and Environmental Sciences, Ludwig-Maximilians Universität München, Munich, Germany \newline
(2) Scripps Institution of Oceanography, UC San Diego, La Jolla, CA, USA \\

Alice-Agnes Gabriel is also affiliated with (1)

Corresponding author:
Nico Schliwa, nico.schliwa@geophysik.uni-muenchen.de,
Department of Earth and Environmental Sciences, 
Geophysics, Ludwig-Maximilians-Universität (LMU) München, Theresienstr. 41,
80333 Munich, Germany

%[Total main text (no bib): \wordcountmaintext ]

\newpage
\section*{Abstract}
Dynamic rupture simulations generate synthetic waveforms that account for non-linear source, path, and site complexity.
Here, we analyze millions of spatially dense waveforms from 3D dynamic rupture simulations in a novel way to illuminate the spectral fingerprints of earthquake physics.
We define a Brune-type \textit{equivalent near-field corner frequency} ($f_c$) to analyze the spatial variability of ground motion spectra and unravel their link to source complexity.
We first investigate a simple 3D strike-slip setup, including an asperity and a barrier, and illustrate basic relations between source properties and $f_c$ variations. 
Next, we analyze $>$ 13,000,000 synthetic near-field strong motion waveforms  generated in three high-resolution dynamic rupture simulations of real earthquakes, namely, the $M_w$ 7.1 2019 Ridgecrest mainshock, the $M_w$ 6.4 Searles Valley foreshock, and the $M_w$ 7.3 1992 Landers earthquake. All scenarios consider 3D fault geometries, topography, off-fault plasticity, viscoelastic attenuation, 3D velocity structure, and resolve frequencies up to 1-2\,Hz. 

Our analysis reveals pronounced and localized patterns of elevated $f_c$, specifically in the vertical components.
We validate such $f_c$ variability in observed near-fault spectra.
Using isochrone analysis, we identify the complex dynamic mechanisms that explain the rays of elevated $f_c$ and cause unexpectedly impulsive, localized, vertical ground motions.
While the vertical high frequencies are also associated with path effects, rupture directivity, and coalescence of multiple rupture fronts, we show that they are dominantly caused by rake-rotated surface-breaking rupture fronts that decelerate due to fault heterogeneities or geometric complexity.
Our findings highlight the potential of spatially dense ground motion observations for furthering our understanding of earthquake physics directly from near-field data.
Observed near-field $f_c$ variability may inform on directivity, surface rupture, and slip segmentation. Physics-based models can identify ``what to look for'', for example, in the potentially vast amount of near-field large array or Distributed Acoustic Sensing (DAS) data.

\section*{Introduction} \label{sec:Introduction}
The advances of seismic array analysis (e.g., \citealp{RostThomas2002}; \citealp{DoughertyEtAl2019}; \citealp{ArrowsmithEtAl2022}), the rise of Distributed Acoustic Sensing (DAS) \citep[e.g.,][]{Zhan2019} and detailed displacement analysis using high-rate GNSS networks \citep[e.g.,][]{MadariagaEtAl2019, PaziewskiEtAl2020} highlight the potential of dense ground motion observations. 
Near-field recordings of well-instrumented earthquakes \citep{Siehetal1993, ChungShin1999, LangbeinEtAl2005, RossEtAl2019} have revealed large variability of ground motions which may originate from local site, path, and source effects \citep{ShakalEtAL2006, RippergerEtAl2008}.
For example, \citet{OlsenEtAl2008} report ``star burst patterns'' of increased ground motion peak values radiating out from the San Andreas fault where a dynamic rupture pulse changes abruptly in either speed, direction, or shape. 

Concurrently, numerical forward simulations of earthquakes, combining realistic modeling of seismic sources and wave propagation, have advanced tremendously over the last decades and can provide realistic and spatially dense ground motion synthetics.
Earthquakes on prescribed finite-fault geometries are modeled using kinematic (\citealp{Ben-Menahem1962}; \citealp{Haskell1964}) or dynamic (\citealp{Madariaga1976}; \citealp{Andrews1976}) approaches.
Kinematic models prescribe co-seismic slip evolution and are computationally more efficient but do not guarantee a physically consistent source description (e.g., \citealp{TintiEtAl2005b}).
Dynamic rupture models provide mechanically viable descriptions of how faults yield and slide based on laboratory-derived friction laws and can provide physics-based correlations among macroscopic earthquake rupture parameters, such as slip rate and rupture time \citep{GuatteriEtAl2004, SchmedesEtAl2010, SavranOlsen2020, VyasEtAl2023}.
High-performance computing allows deterministic modeling of the broadband seismic wavefield in the near-field of kinematic and dynamic earthquake models across the bandwidth of relevance for earthquake engineering \citep[e.g.,][]{Heineckeetal2014, WithersEtAl2018, RodgersEtAl2019, TaufiqurrahmanEtAl2022}. A high degree of realism of 3D physics-based forward simulations can be achieved by integrating observational datasets, such as high-resolution velocity structure and topography \citep[e.g.,][]{Smalletal2017, PitarkaEtAl2021}, as well as physically relevant mechanisms beyond elasticity and simple source geometries, such as fault zone plasticity, viscoelastic attenuation, and fault roughness and segmentation \citep[e.g.,][]{ShiDay2013, RotenEtAl2017, Wollherretal2019, BoLiEtAl2023, TaufiqEtAl2023}.

Various connections between earthquake source complexity and the variability of observed and modeled strong ground motions have been identified.
Surface-breaking of dynamic rupture can cause large fault-parallel ground velocity pulses \citep{KanekoGoro2022}.
High-frequency and high-intensity radiation that dominates ground acceleration can be generated by abrupt changes in rupture velocity, heterogeneity of slip or slip rate, or variations in fault geometry \citep{Madariaga1977, Madariaga1983, SpudichCranswick1984, HartzellEtAl1996, MadariagaEtAl2006, ShiDay2013}. 
Small-scale ruptures in laboratory experiments and large earthquakes analyzed using back-projection emit high‐frequency radiation close to the rupture tip \citep[e.g.,][]{MartyEtAl2019,LiEtAl2022}.
\citet{PulidoDalguer2009} analyzed the high-frequency radiation of large-stress drop regions (``asperities'') and areas with a larger strength excess than their surroundings (``barriers'').
They found that most of the high-frequency radiation during the 2000 Tottori earthquake originated from only 20\% of the total asperity area, thereby highlighting the local character of the generation mechanism(s).
Envelope inversions of ground accelerations show that high-frequency waves mainly radiate near the periphery of the fault plane or at the boundary of large slip areas (\citealp{ZengEtAl1993}; \citealp{KakehiIrikura1996}). 
Using a hybrid back-projection method, \citet{OkuwakiEtAl2014} observed that strong high-frequency radiation precedes the large asperity rupture of the 2010 $M_w$ 8.8 Chile earthquake. 

Recent observational data analyses imply that high-frequency radiation may strongly correlate with fault trace ``misalignment'' \citep{ChuEtAl2021}, i.e., geometric fault complexity.
\citet{Adda-BediaMadariaga2008} showed theoretically that a rupture front turning at a fault kink generates a burst of high-frequency radiation aligned with a jump in particle velocity, which has also been observed in 3D dynamic and kinematic rupture simulations \citep{OglesbyMai2012, LiEtAl2022}.
\citet{ZhangGe2017} reported peaks in high- and low-frequency seismic energy release at a stepover rupture during the 2014 $M_w$ 7.0 Yutian earthquake. 
While rupture directivity affects dominantly low and intermediate frequency bands, it can cause most of the seismic energy from a finite rupture to arrive as a single large pulse and may increase the components' average high-frequency radiation in a magnitude-dependent frequency band \citep{SommervilleEtAl1997, KaneEtAl2013}.
The strength of directivity effects depends on the ratio of the mean rupture velocity to wave-propagation velocity \citep{BooreJoyner1978}.
3D dynamic rupture simulations have shown that off-fault plasticity causes near-fault peak ground velocities to saturate and increases the dominant period of such a directivity pulse \citep{WangDay2020}.

However, identifying and physically interpreting observable near-field ground motion complexities remains challenging.
Theoretical source models often assume simple, for example, circular, source geometries and a constant sub-shear/sub-Rayleigh rupture speed.
Dynamic rupture simulations have demonstrated that this is rarely the case for large earthquakes (e.g., \citealp{Ulrichetal2019}; \citealp{UlrichEtAl2019b}; \citealp{HarrisEtAl2021}; \citealp{YuEtAl2022}).
Additionally, even for the best-recorded large earthquakes, observations often miss the spatial resolution required to uniquely relate ground motion variability to source complexity.

Here, we aim to systematically relate the spectral properties of synthetic strong ground motions from 3D dynamic rupture earthquake scenarios to physics-based source complexity.
We analyze millions of synthetic waveforms from dynamic rupture simulations to illuminate the spectral fingerprints of earthquake source mechanisms in the near-field.
We define a spatially variable Brune-type \citep{Brune1970} corner frequency $f_c$ as a scalar proxy of a waveform's relative frequency content to analyze its variability in the vicinity of rupturing fault systems and associate it with different aspects of source complexity. We term $f_c$ as the \textit{equivalent near-field corner frequency} to avoid any confusion with far-field corner frequency analysis.
We analyze four 3D dynamic rupture models of increasing complexity for which we generate spatially high-resolution ground motion synthetic seismograms.
We identify distinct spatial patterns in $f_c$ that are associated with fault geometry complexity, rupture directivity, surface rupture, or variable slip distribution.
We use isochrone theory (\citealp{SpudichCranswick1984}; \citealp{BernardMadariaga1984}) to locate sources of high-frequency radiation and interpret our results.

\section*{Methods} \label{sec:Methods}

\subsection*{Equivalent near-field corner frequency}  \label{sec:methods_fc}
The average corner frequency of far-field source spectra can be used to estimate an event's stress drop (e.g., \citealp{ShearerEtal2006}, \citealp{AllmannShearer2009}, \citealp{Abercrombie2021}), which requires a theoretical model of the source. 
The classical \citet{Brune1970} model describes the displacement amplitude spectrum $A(f)$ of far-field body waves as follows:
\begin{equation}
A(f) = \frac{\Omega_0}{1+(f/f_c)^2}\,,
\label{eq:Brune_corner_frequency}
\end{equation} 
where $\Omega_0$ is the amplitude of the lowest frequency of the spectrum, $f_c$ is the corner frequency, and $f$ is a well-defined frequency band. 
The Brune-type spectrum is flat at low frequencies, with $\Omega_0$ proportional to the seismic moment $M_0$, and it has an $\omega^{-2}$ fall-off rate at high frequencies.
The corner frequency $f_c$ marks the transition between the two parts of the spectrum. 

Here, we aim to analyze the relative spatial variability of simulated spectra in the near-field in distinction to inferring source properties, such as source dimensions or stress drop. We adapt the classical \citet{Brune1970} model (\hyperref[eq:Brune_corner_frequency]{Equation \ref{eq:Brune_corner_frequency}}) to determine near-field spectral corner frequencies. 
We acknowledge that our application violates some of \citet{Brune1970}'s underlying theoretical assumptions: our simulated waveforms include non-negligible near-field terms, the effects of topography and 3D velocity structure, and a clear separation between P- and S-wave spectra is mostly impossible because the event durations are longer than the arrival time differences (e.g., \citealp{MadariagaEtAl2019}).
Thus, we term the inferred spectral waveform property the \textit{equivalent near-field corner frequency ($f_c$)}.

First, we rotate the horizontal components of all synthetic seismograms into radial and transverse components with respect to the absolute slip centroid of the respective dynamic rupture model.
We then apply a tapered body wave window to each time series to mitigate the impact of later arriving surface waves.
The body wave window effect is generally small, with the exception of the Searles Valley dynamic rupture simulation, which is discussed in the \hyperref[sec:Results]{Results}.
The length of the body wave window is chosen as the respective rupture duration plus an S-wave delay specifically calculated for each virtual station based on its distance to the slip centroid.   

Next, we Fourier transform the velocity waveforms and integrate the spectra by division with $i\omega$. The order of these operations is important because computing the Fourier transform of an already integrated displacement time series, which potentially contains static displacement and  is thus not periodic, can lead to spectra that are contaminated at all frequencies \citep{MadariagaEtAl2019}. In the next step, we resample the spectrum to equally spaced sampling points up to the numerically resolved maximum frequency of each simulation (see \hyperref[sec:Models]{3D dynamic rupture models} in \hyperref[sec:Appendices]{Appendices}). This equal spacing is necessary to maintain the misfit to be independent of the sampling density in the frequency domain. We note that \citet{AllmannShearer2009} addressed this problem differently by using weights to account for sampling density. 

We solve \hyperref[eq:Brune_corner_frequency]{Equation \ref{eq:Brune_corner_frequency}} for all possible $f_c$ values in 0.005\,Hz steps between the inverse of the body wave window (always $<$ 0.1\,Hz) and 1.0\,Hz, and evaluate the misfits between the simulated and analytical spectra.
We define the equivalent near-field corner frequency as the value of all possible $f_c$ values, which leads to the smallest misfit.
Theoretically, $\Omega_0$ in \hyperref[eq:Brune_corner_frequency]{Equation \ref{eq:Brune_corner_frequency}} is given by the amplitude of the lowest frequency of the respective spectrum. However, the fit can be generally improved by considering a mean amplitude value of the low-frequency part (e.g., \citealp{Trugman2020}). Here, we choose the mean value of the lowest frequency up to the respective $f_c$, which renders our approach robust for spectra that contain static displacements.

We use the Spectral Seismological misfit approach of \citet{Karimzadeh2018} for corner frequency picking:
\begin{equation}
    \mathrm{Misfit_{\,SS}} = \frac{1}{n_f} \displaystyle\sum_{i=1}^{n} |log \frac{A(f_i)}{A_{Brune}(f_i)}|\,,
    \label{eq:Spectral_seismological_misfit}
\end{equation}
where $f_i$ are the discrete sample points of the spectra, and $n_f$ is the absolute number of sample points. $A(f_i)$ are the spectral amplitudes of the synthetic waveforms and $A_{Brune}(f_i)$ are the amplitudes of the respective Brune-type spectra (\hyperref[eq:Brune_corner_frequency]{Equation \ref{eq:Brune_corner_frequency}}). This approach has two major benefits compared to a root-mean-square misfit. It is based on relative differences and, therefore, is independent of the examined absolute amplitudes. Additionally, its logarithmic scaling prevents overweighting of outliers in strongly oscillating spectra. 

We compute $f_c$ using parallelized Python code and exploiting efficient NumPy tensor operations \citep{HarrisEtAl2020}. We openly provide our code (see \hyperref[sec:DataAndResources]{Data and Resources}) that automatically facilitates the loading of raw waveform data, preprocessing, and calculation of $f_c$. For example, our script requires approximately 23 minutes to process 3,000,000 waveforms of the Landers earthquake dynamic rupture scenario using 30 processes.
Input data loading is also parallelized but does not optimally scale and uses approximately 50 \%  of the computational runtime. We perform the computations on an AMD EPYC 7662 64-Core processor with a 2\,GHz clock speed.

\subsection*{Isochrone theory}
We use isochrone theory (\citealp{SpudichFrazer1984}; \citealp{BernardMadariaga1984}) to interpret the equivalent near-field corner frequency distributions.
Iso\-chrone theory assumes that close to large earthquakes, most high-frequency ground motions are caused by direct P- and S-waves generated at the rupture front (\citealp{Madariaga1983}; \citealp{SpudichCranswick1984}; \citealp{MartyEtAl2019}).
Under this assumption, high-frequency ground motions recorded at a station can be derived from a series of line integrals instead of using the full surface integral of the representation theorem.

In the framework of isochrone theory, the integration path for each time step consists only of points on the fault associated with the high-frequency radiation that arrives at the observer at the respective time step.
These lines are called isochrones and are contour lines of the sum of the rupture times and travel times to the respective station.
Rupture time is defined as the time at which the absolute slip rate at a point on the fault exceeds 0.001\,m/s.
The ground velocities can be directly related to the isochrone velocity $c$, which is proportional to the isochrone spacing as
\begin{equation}
    c(x, z, \underline{r}) = |\nabla_s t(x,z,\underline{r})|^{-1}\,,
    \label{eq:isochrone_velocity}
\end{equation}
and $\nabla_s$ is the surface gradient with respect to the fault coordinates ($x$,$z$), $t$ is the isochrone time (the sum of the rupture time and travel time), and $\underline{r}$ are the coordinates of the station. Points on the fault where $c$ is singular radiate particularly high frequencies. A prominent example is supershear rupture and its associated S-wave Mach cones \citep{SpudichFrazer1984}. Spatial variations in the slip velocity and temporal variations in the isochrone velocity can cause comparable ground accelerations.
Seismic directivity decreases isochrone spacing, thereby increasing isochrone velocity \citep{SpudichChiou2008}. Thus, isochrone analysis inherently captures the contributions of near-field directivity effects.

To use isochrone theory to analyze complex dynamic rupture scenarios, here, we mostly use the peak slip rate time instead of the rupture (initiation) time.
Peak slip rate time is also associated with the rupture front but is often smoother and less prone to ambiguity, for example, due to multiple rupture (and healing) fronts.
We only assign peak slip rate times to points on the fault where the peak slip rate exceeds 0.05\,m/s for TPV5 and 0.1\,m/s for the Ridgecrest and Landers dynamic rupture models.
In the following, we mostly show peak slip rate times inferred from the dip-slip components to separate the vertical slip from the strike-parallel slip and to isolate the effects of rake-rotated rupture fronts.

\section*{Results} \label{sec:Results}
We analyze the waveforms generated in 3D dynamic rupture simulations of a simple community benchmark setup and three large-scale scenarios of real earthquakes. All four dynamic rupture scenarios are detailed in the \hyperref[sec:Appendices]{Appendices}.

\subsection*{TPV5 3D dynamic rupture community benchmark}
\begin{figure}[!htb] 
\centering
\includegraphics[scale=0.38]{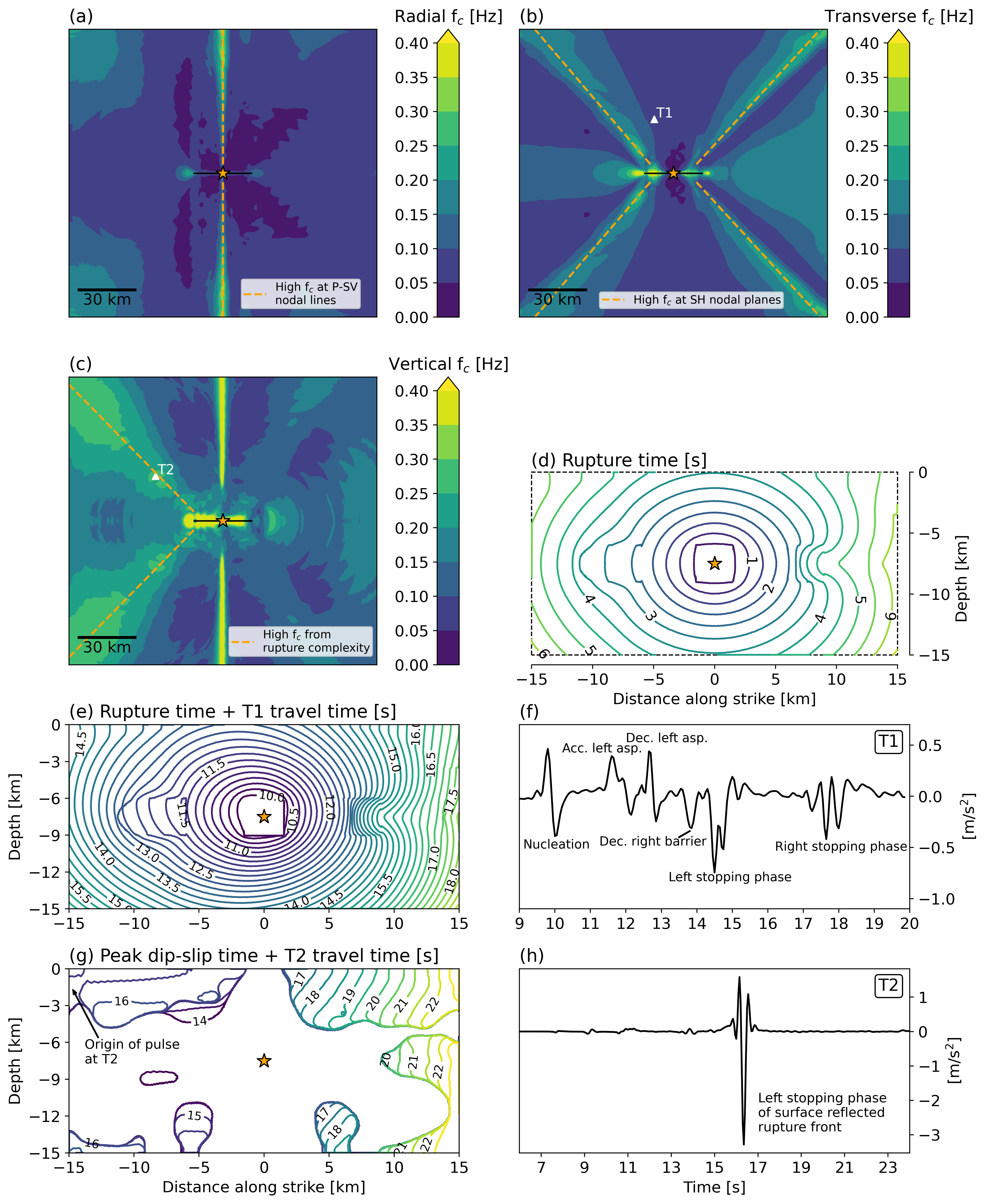}
\caption{Equivalent near-field corner frequency ($f_c$) distribution of the (a) radial, (b) transverse, and (c) vertical components of synthetic seismograms recorded at $\sim$900,000 virtual seismic stations in map view. The seismograms are generated in a bilateral strike-slip 3D dynamic rupture model including an asperity and a barrier embedded in a homogeneous elastic half-space (the TPV5 SCEC/USGS community benchmark \citep{HarrisEtAl2009}). Black lines indicate the fault trace, the star marks the hypocenter, and white triangles are stations that are analyzed in subplots e-h. Orange lines mark different high-$f_c$ features. (d) Side-view of the fault plane with rupture front contours in 0.5\,s intervals. (e) Isochrone contours of station T1 in 0.5\,s intervals. (f) Transverse ground accelerations at station T1. Comparison to isochrones allows associating pronounced high-amplitude signals with different stages of 3D dynamic rupture propagation. Acc. = Acceleration; dec. = Deceleration; asp. = Asperity. (g)  Peak dip-slip isochrone contours of station T2 in 0.5\,s intervals. (h) Vertical ground accelerations at station T2.}
\label{fig:tpv5_fc_4x2.png}
\end{figure}

We demonstrate the relationship between the source properties, high-frequency radiation, and equivalent near-field corner frequency variations using the TPV5 USGS/SCEC 3D dynamic rupture community benchmark \citep{HarrisEtAl2009}. 
The benchmark's domain is a homogeneous elastic half-space; therefore, no path or site effects affect our analysis. We model a bilateral strike-slip dynamic rupture passing an asperity and a barrier, regions with elevated or reduced initial shear stresses, which significantly accelerate or decelerate the rupture, respectively (Figure \ref{fig:tpv5_fc_4x2.png}d).
In the following, ``high-frequency'' refers to frequencies higher than the average corner frequency.

Figure \ref{fig:tpv5_fc_4x2.png}e-h shows two pairs of isochrone contours on the fault plane and acceleration waveforms of the respective virtual stations. Isochrones in subplot (e) are calculated by adding the S-wave travel time at station T1 (located at $x = -10$\,km and $y = 30$\,km; see Figure \ref{fig:tpv5_fc_4x2.png}b) to the rupture time. The patch in the middle represents the overstressed nucleation area. The rupture acceleration due to the left asperity causes an increased isochrone spacing to the left side of the hypocenter and vice versa the rupture deceleration due to the right barrier causes a decrease in isochrone spacing to the right side of the hypocenter. Figure \ref{fig:tpv5_fc_4x2.png}f shows the corresponding acceleration time series of the transverse component at T1. Ground accelerations are generally associated with high-frequency radiation and are proportional to changes in the isochrone spacing \citep{SpudichFrazer1984}. Every pronounced high-amplitude signal in the accelerogram can be related to a specific rupture phase by comparing the time step with the respective isochrone. For example, the accelerogram oscillations at 10\,s are caused by the nucleation, the two spikes at 11.5\,s and 12.5\,s represent the acceleration and deceleration at the high-stress asperity, and the strongest pulse at 14.5\,s results from stopping phases when rupture reaches the prescribed left end of the fault.

Figure \ref{fig:tpv5_fc_4x2.png}g shows the peak dip-slip isochrones (the sum of the S-wave travel time and peak dip-slip time) at station T2 (located at $x = -35$\,km and $y = 25$\,km; see Figure \ref{fig:tpv5_fc_4x2.png}c). Those points on the fault with peak dip-slip rates of less than 0.05\,m/s are ignored because their contribution to the radiated waves is negligible. We use a median filter favoring the dominant isochrones to avoid oscillations where different rupture fronts have a comparable peak slip rate. Figure \ref{fig:tpv5_fc_4x2.png}h shows the vertical ground accelerations at station T2, which are dominated by a single spike shortly after 16\,s.
Isochrone analysis reveals that this spike is caused by a phase of dip-slip acceleration and abrupt rupture arrest induced by a surface-reflected rupture front.
The large isochrone spacing (after 16\,s, Figure \ref{fig:tpv5_fc_4x2.png}g) at the upper left side of the fault shows this strongly accelerating dip-slip phase that abruptly stops at the left fault end.
The dip-slip phase originates from rake rotation at the dynamic rupture front, which is larger at shallow depths and is significantly amplified when the rupture breaks the surface. Such shallow rake rotation has been linked to geological features such as slickenlines \citep{KearseKaneko2020} and increased tsunami hazards of strike-slip fault systems \citep{BoLiEtAl2023, Kutschera2023}.

Importantly, the equivalent near-field corner frequency $f_c$ is not a direct measure of the high-frequency content of the modeled seismic wavefield but rather reflects a relative association between high and low frequencies. 
Figure \ref{fig:tpv5_fc_4x2.png}a-c shows the $f_c$ distribution of the radial, transverse, and vertical components of the synthetic seismograms recorded at $\sim$900,000 virtual stations in map-view. 
We observe pronounced variability in $f_c$.
A thin ray of high $f_c$ in fault-normal direction, visible in the radial and vertical components (Figure \ref{fig:tpv5_fc_4x2.png}a,c), reflects the nodal line of the P-SV radiation pattern and is also present in other corner frequency studies of (near) symmetrical ruptures (e.g., \citealp{KanekoShearer2015}; \citealp{WangDay2017}).

Rays of high $f_c$ form in the transverse components (Figure \ref{fig:tpv5_fc_4x2.png}b) at an angle of approximately 45$\degree$ to the rupture propagation direction.
These reflect the nodal planes of a strike-slip SH radiation pattern centered at the hypocenter.
The radiation pattern affects low frequencies stronger than high frequencies (e.g., \citealp{TakemuraEtAl2009}), which leads to a lack of low-frequency energy at the nodal planes and thus locally increases the measured $f_c$.

We observe two ray-like high-$f_c$ patterns at approximately the same 45$\degree$ angle to the fault trace in the vertical components (Figure \ref{fig:tpv5_fc_4x2.png}c), for example, where station T2 is placed.
We link these to the high-amplitude lobes of a strike-slip P-SV radiation pattern in the rupture-forward direction of a vertical high-frequency pulse caused by a rapidly accelerating and decelerating phase of dip-slip during surface-breaking rupture, as shown in the isochrones in Figure \ref{fig:tpv5_fc_4x2.png}g.
We quantify the rake rotation related to shallow dip-slip to be only up to 10$\degree$ (Figure \ref{fig:rake_rotation_figure}a). Thus, the overall radiation is dominated by the strike-slip radiation pattern.
Directivity effects additionally sharpen the vertical ground motion pulse and contribute to the high $f_c$.
The vertical-component high-$f_c$ bands are of considerably lower amplitudes on the right side of the model domain. This is caused by the differences in bending of the rupture front due to either high- or low-shear stress patches (Figure \ref{fig:tpv5_fc_4x2.png}d). 
Convex bending due to the submerged left, high shear stress asperity leads to more abrupt decelerating of the surface-reflected rupture front, generating more high-frequency radiation.

\subsection*{2019 Ridgecrest sequence} \label{subsec:ridgecrest}

We analyze the relationship between the equivalent near-field corner frequency $f_c$, isochrones, and source complexity in \citet{TaufiqEtAl2023}'s 3D dynamic rupture scenario for the 2019 $M_w$ 7.1 Ridgecrest mainshock.
We also analyze the $f_c$ distribution of their Searles Valley foreshock dynamic rupture model, which reveals additional path effects. Figure \ref{fig:ridgecrest_overview_plot}a-c provide an overview of both dynamic rupture models and the \hyperref[sec:Appendices]{Appendices} include a detailed description.

The Ridgecrest mainshock dynamic rupture scenario ruptures primarily along an NW-SE-trending continuous fault (F3 in Figure \ref{fig:ridgecrest_overview_plot}c).
The rupture starts as a bilateral crack that expands away from the hypocenter. 
After 5\,s rupture time, it terminates to the north due to a locally lower prestress and a less optimal fault orientation.
The southward rupture can not break through the conjugate F2-F3 intersection at shallow depths due to the stress shadow caused by the foreshock dynamic rupture scenario \citep{TaufiqEtAl2023}.
Only deep decelerated slip "tunnels" the intersection and regrows as a pulse that ruptures again to the surface and to the southeastern end of F3.

Figure \ref{fig:RC_e2_bw_fc_2x2} shows the $f_c$ distribution of the three components of the Ridgecrest mainshock dynamic rupture simulation, complemented by a map of the regional topography incorporated into the model and the fault system surface traces.
For each component, the equivalent near-field corner frequencies are computed at approximately 1,800,000 virtual seismic stations with a spacing of $\sim$500\,m. 
The vertical components exhibit particularly high spatial variability in the inferred $f_c$.

We find that the directivity effects associated with the bilateral rupture lead to an elevated $f_c$ at both ends of the main fault in the radial and vertical components (Figure \ref{fig:RC_e2_bw_fc_2x2}a,c and red dashed lines in Figure \ref{fig:RC_e2_fc_map_3x1}d). 
The corner frequency variability of the transverse component (Figure \ref{fig:RC_e2_bw_fc_2x2}b) is smaller than that of the other components and resembles to first-order a strike-slip radiation pattern: 
$f_c$ is higher close to the nodal planes and lower where the wavefield amplitudes are expected to be the largest. 

A gap between two high-$f_c$ rays in the vertical components (orange lines in Figure \ref{fig:RC_e2_fc_map_3x1}d) is related to a major rupture complexity of the mainshock, the "tunneling" dynamics at the intersection with the orthogonal fault F2 (see Figure \ref{fig:ridgecrest_overview_plot}c).
We compare the modeled acceleration waveforms at synthetic stations located within the elevated-$f_c$ regions (stations R1 and R3) with a station located in the gap in between the high-$f_c$ rays (station R2) in Figure \ref{fig:RC_e2_fc_map_3x1}a-c. The waveforms show that regions with high-$f_c$ values are associated with a high-frequency acceleration pulse that is absent in the R2 accelerogram. 

Figure \ref{fig:RC_e2_fc_map_3x1}e,f show the corresponding peak dip-slip isochrones at high-$f_c$ stations R1 and R3, which we use to identify the origin of the acceleration pulses.
The acceleration pulse at R1 occurs shortly after 17\,s simulation time.
This 17\,s pulse originates from an ``isochrone jump'' at the intersection with the orthogonal fault F2 close to the surface.
This isochrone jump is caused by the abrupt stopping of the rupture at the intersection at shallow depth and delayed activation of the fault area behind the intersection.
The complex shape of the peak dip-slip isochrones in the shallow area before the fault intersection (Figure \ref{fig:RC_e2_fc_map_3x1}e) is caused by a secondary surface-reflected rupture front, which involves a pronounced rake rotation (Figure \ref{fig:rake_rotation_figure}b).
The complex dynamics of surface rupture, rake rotation, and abrupt rupture arrest at the geometric barrier formed by the fault intersection conjointly generate pronounced high-frequency radiation in the vertical component.
The high-$f_c$ ray, in which R1 is located (Figure \ref{fig:RC_e2_fc_map_3x1}d), reflects a high-amplitude lobe in the rupture-forward direction of a P-SV radiation pattern of this high-frequency dynamics.

The same mechanism of surface dynamic rupture, rake rotation, and arrest explains the high acceleration pulse recorded at R3.  
This pulse occurs at 20\,s which coincides with the surface-breaking and rake-rotated rupture front stopping abruptly at the southern end of the fault system.
Its amplitude is higher since this station is closer to the fault, the directivity effect is stronger, and rupture deceleration may appear as more abrupt.
The isochrones show the first upgoing rupture front, whereas the vertical acceleration pulses are associated with the deceleration of the surface-reflected rupture front.
This leads to a timing discrepancy of approximately 1\,s between the high-frequency pulse in the accelerogram and the rupture stopping indicated by the isochrones.

The spectra of the observed near-field vertical ground motions show similar spatial variability in $f_c$ as the dynamic rupture model of the 2019 Ridgecrest mainshock. 
In Figure \ref{fig:RC_e2_fc_map_3x1}d, we show the vertical $f_c$ values of the observed spectra at 17 near-field stations.
The observed $f_c$ values depend on azimuth and mostly resemble the synthetic $f_c$ map.
Six stations are located in rupture forward direction at the northern end of the fault system, and $f_c$ values at four of these stations agree with our modeled values while $f_c$ at two stations is lower than in our model.
We may speculate, that a slightly different location of rupture arrest to the North or unmodeled site effects cause this discrepancy.
The station with the highest observed $f_c$ is located at the southern end of the fault system where the strongest directivity is expected, which agrees with our model.

The dynamic rupture model of the Searles Valley foreshock initiates close to the F1-F2 fault intersection (Figure \ref{fig:ridgecrest_overview_plot}c).
Right-lateral slip on F1 spontaneously ceases without reaching the surface, which agrees with observations \citep{LiuEtAl2019}.
The deep rupture on F1 activates the conjugate, critically prestressed left-lateral fault F2.
F2 ruptures entirely to its southwestern end, accumulating most of the event's slip and breaking the surface over its full length.

Although both events, the mainshock, and the foreshock, are rupturing the same fault system their vertical $f_c$ distributions differ vastly. This illustrates that $f_c$ is dominated by source effects. 
In Figure \ref{fig:RC_e1_fc_map_3x1}d we observe a wide $f_c$-shadow zone, an area of smaller than average $f_c$, in the rupture backward-directivity direction (NE) which reflects the dominantly unilateral nature of the foreshock rupture. 
A thin ray of elevated $f_c$ within this shadow zone emits from the small non-surface rupturing slip patch at the NW trending F1 (Figure \ref{fig:ridgecrest_overview_plot}c). 
Our results show high-$f_c$ structures (highlighted with dashed orange lines in Figure \ref{fig:RC_e1_fc_map_3x1}d) pointing away from the southwestern part of the primary fault F2, where the rupture breaks the surface. 
A gap in these high-$f_c$ rays coincides with a small kink of the fault trace. 

A pronounced feature in the vertical $f_c$ distribution of the Searles Valley foreshock is a curved-path high-$f_c$ ray (dashed red line in Figure \ref{fig:RC_e1_fc_map_3x1}d), which changes direction from west-northwest to northeast.
Its bent structure is caused by deflection at the strong velocity contrast along the southern Sierra Nevada mountain range (see Figure \ref{fig:RC_e2_bw_fc_2x2}d).
An animation of the 3D wavefield (see \hyperref[sec:DataAndResources]{Data and Resources}) illustrates these locally strong path effects. 

Seismic energy is directed in unexpected directions and significantly prolongs the observed shaking duration at several stations, e.g., at station WMF.
Figure \ref{fig:RC_e1_fc_map_3x1}a-c shows synthetic and observed waveforms at station WMF, which are lowpass-filtered at 0.5\,Hz. 
In this frequency range, the synthetic ground motions of the directly arriving wave packet agree well with the observations (before 40\,s). 
For the second wave packet, which is associated with the deflection at the mountain range, there are significant differences between the synthetic and observed waveforms. 
The synthetic wave packet arrives earlier, has a shorter duration, and contains single high-frequent spikes in the EW and Z components. 
These differences may be explained by a lack of small-scale subsurface heterogeneity in the 3D velocity model used in the dynamic rupture simulation.
Additional high-frequency wave scattering may prolong coda-shaking \citep[e.g.,][]{ImperatoriMai2012, TakemuraEtAl2015, TaufiqurrahmanEtAl2022}. 
The travel time difference may be caused by underestimating the velocity reduction of a sedimentary basin between the mountain range and the fault system in the used 3D velocity model (CVM-S4.26; \citep{Leeetal2014}). 
In order to not omit late-arriving deflected waves, we derive $f_c$ values in Figure \ref{fig:RC_e1_fc_map_3x1}d from the full-time series of the simulation (100\,s duration after the nucleation) without picking a body wave window.
The same plot with selecting a body wave window is shown in Figure S2.
The body wave window removes the curved high-$f_c$ ray, but otherwise, the $f_c$ distribution is nearly identical. 

\subsection*{1992 Landers earthquake}

The dynamic rupture model of the 1992 Landers earthquake by \citet{Wollherretal2019} ruptures across five fault segments (see Figure \ref{fig:landers_overview_plot}, and for an animation see \hyperref[sec:DataAndResources]{Data and Resources}).
Rupture nucleates at the southern part of the Johnson Valley fault (JVF) and propagates northward.
After 4\,s, the rupture migrates to the Kickapoo fault (KF) by direct branching. 
The Kickapoo fault connects the rupture from the JVF to the Homestead Valley fault (HVF).
The rupture nearly stops at a fault bend at the HVF but then reinitiates and breaks up to its northern extent.
While rupturing the HVF, a second rupture front branches to the Emmerson fault (EF).
At around 15.5\,s, a part of the EF is also activated by dynamic triggering from waves generated at the northern part of the HVF.
Multiple rupture fronts (including back-propagating rupture fronts) form when the slower rupture front from direct branching reaches the part of the EF that was dynamically triggered.
The backward propagating rupture dynamically reactivates parts of the HVF and the KF.
After 22.3\,s, a completely separate subevent on the Camp Rock fault (CRF) is dynamically triggered by the superimposed waves generated at the EF and the northern part of the HVF.
The rupture completely stops after 30\,s.

Figure \ref{fig:Landers_fc_2x2}a-c shows the equivalent near-field corner frequency $f_c$ distributions of the Landers dynamic rupture simulation computed at $\sim$1,000,000 virtual seismic stations.
The vertical components, as well as parts of the radial components, exhibit considerable spatial variability in the inferred $f_c$, whereas the spatial variability of $f_c$ of the transverse components is lower.
The vertical component shows a complex pattern of localized rays of increased $f_c$, pointing away from the fault trace.
We observe a correlation between the distribution of $f_c$ in the vertical and radial components which we interpret as an indication that the P-SV wave modes are responsible for the observed rays.

Sharp changes of $f_c$ outline several low-velocity sedimentary basins, such as the Salton Sea basin in the South and the San Bernardino and Los Angeles basins in the South-West. 
Low-velocity sediment basins lead to sharp corner frequency increases of the P-SV mode at their edges. While this is a plausible mechanism, we here clip the color map at sedimentary basins and close to the fault, where static displacement and an inaccurate component separation due to finite-fault effects distort the corner frequency determination and we omit these regions in our interpretation. 

We highlight the rays of high $f_c$ in the vertical components of the Landers model as dashed lines in Figure \ref{fig:Landers_fc_map_3x1}d.
These rays mostly take off at an angle close to 45$\degree$ to the rupture forward direction, and they trace the curvature of the segmented fault system.
South of the epicenter, $f_c$ is generally lower, although this is overprinted by a local increase due to the Salton Sea basin. 
Decreased southern $f_c$ is caused by the directivity effects of the north-northwestwards-directed unilateral rupture, which prolongs shaking in the backward direction and mitigates the high-frequency content.

Three seismic stations (Fort Irwin, Barstow, and Yermo), which recorded near-field waveforms of the 1992 Landers event, are located on or between our modeled rays of elevated corner frequencies.
Comparing the observed spectra with our synthetic data reveals an excellent fit between 0.1 and 1.0\,Hz (Figure \ref{fig:Landers_fc_map_3x1}a-c).
Station Fort Irwin is located between two high-$f_c$ rays and has a lower $f_c$ than the spectra of the stations Barstow and Yermo, which are located on high-$f_c$ rays.  
We use isochrone theory to identify the source of the elevated $f_c$ rays in the vertical components.
The rupture process of the Landers model is complicated and often involves multiple rupture fronts and reactivation of slip, while the isochrones can capture only a single phase of slip (see the animation of the rupture process, \hyperref[sec:DataAndResources]{Data and Resources}).
Figure \ref{fig:Landers_isochrones}a,c,e shows the synthetic ground accelerations at three selected stations: Yermo and two virtual stations L1 and L2 (Figure \ref{fig:Landers_fc_map_3x1}d). 
As expected from previous observational analysis \citep{CampbellBozorgnia1994}, directivity effects strongly affect waveforms recorded at Yermo, which is located in the average strike direction of the unilateral rupture.
Therefore, Yermo's waveforms overall have a considerable amount of high-frequency shaking and do not exhibit a single spike that is solely responsible for the high $f_c$.
We observe, however, that a dynamically triggered subevent at the Camp Rock fault causes the highest waveform amplitudes at about 36\,s (Figure \ref{fig:Landers_isochrones}a,b).
This subevent initiates at 8\,km depth, breaks the surface, and terminates during a short period of time, which superimposes the overall source spectrum with a source spectrum of a lower magnitude subevent corresponding to a higher $f_c$.
The same subevent causes a strong signal at about 40\,s at L1.

\begin{figure}[!htbp] 
\centering
\includegraphics[scale=0.55]{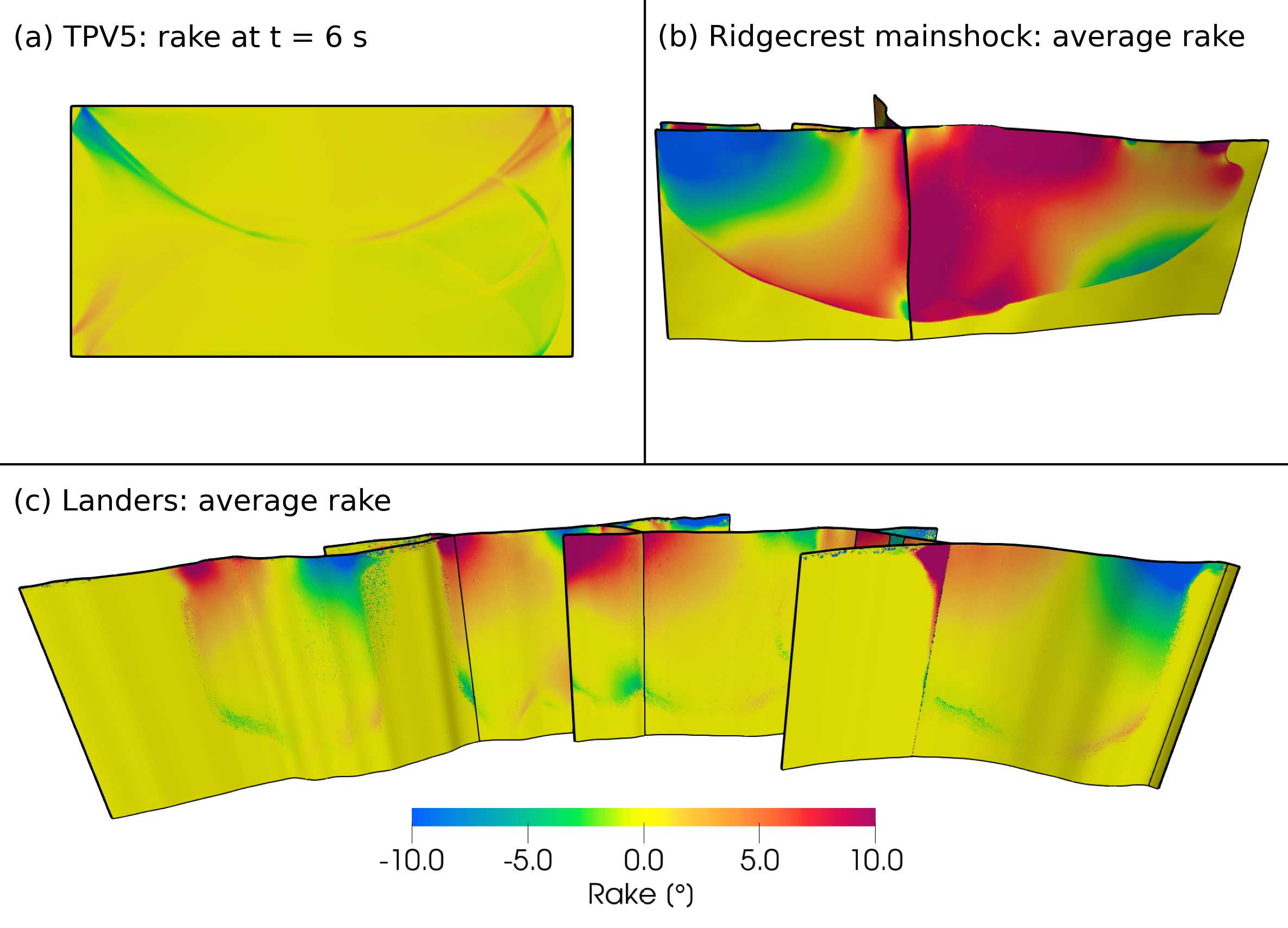}
\caption{(a) Rake at t = 6\,s of the TPV5 dynamic rupture model \citep{HarrisEtAl2009}. (b) Average rake of the 2019 $M_w$ 7.1 Ridgecrest mainshock dynamic rupture model \citep{TaufiqEtAl2023}. (c) Average rake of the 1992 $M_w$ 7.3 Landers dynamic rupture model \citep{Wollherretal2019}.}
\label{fig:rake_rotation_figure}
\end{figure}

The accelerogram at virtual station L1 is dominated by a pulse starting at about 29\,s.
The isochrones show that the source of the high-frequency radiation recorded at L1 is located either at the top or at the bottom of the rupture zone of the EF. 
The animation of the rupture shows that a dynamically activated back-propagating rupture front coalescences with a forward-propagating rupture front which causes high-amplitude up- and down-dip propagating rupture fronts (see Figure S3c,d).
The down-dip propagating front arrests when it reaches the bottom of the seismogenic zone, and the up-dip propagating front breaks the surface until it stops abruptly at the kink to the CRF.
This shallow surface-breaking rupture is dominant in generating the vertical acceleration pulse at L1, as it involves an average rake rotation of up to 10$\degree$ (Figure \ref{fig:rake_rotation_figure}c), which is not observed at the bottom of the rupture zone.

The accelerogram of the virtual station L2 contains two strong high-frequency pulses, the first starting at about 16\,s and the second starting at 19\,s.
The isochrones show that the first high-frequency pulse coincides with the surface rupture at the kink between the JVF and the KF; therefore, the source might be a so-called kink wave \citep{Adda-BediaMadariaga2008} from the rake-rotated rupture front propagating along the surface.
However, another rupture front propagates up-dip along the kink simultaneously (Figure S3a).
This coalescence of differently directed rupture fronts likely also contributes to the modeled high-frequency radiation.
A similar mechanism, a coalescence of an up-dip propagating rupture front and an along-strike propagating rupture front at the fault kink between the KF and the HVF segments generates the second high-frequency pulse (Figure S3b).
There is a small discrepancy between the isochrone timing and the high-amplitude spike in the waveforms because the isochrones capture the rupture front that propagates up-dip along the kink, while the along-strike propagating rupture front arrives later but still interacts with the former.
We observe large and localized shallow rake rotation at both kinks (Figure \ref{fig:rake_rotation_figure}c).

\section*{Discussion} \label{sec:Discussion}
In our analysis of four dynamic rupture scenarios of large strike-slip earthquakes with varying source complexity, we find localized rays of elevated $f_c$ in the vertical components of each event.
Albeit path and site effects, this variability in equivalent near-field corner frequencies is dominated by source effects. 
The large vertical high-frequency radiation is caused by complex dynamic mechanisms including shallow dip-slip generated by a rake-rotation when the rupture breaks the surface and strong stopping phases due to rupture complexity, e.g., at a barrier or a fault kink.
The high-amplitude lobes of the P-SV radiation pattern and the directivity effect shape the rays to form a 45$\degree$ angle to the rupture forward direction.

Our results imply that high-$f_c$ rays correlate with certain characteristic rupture dynamics.
The rays often point to locations where dynamic rupture decelerates suddenly, specifically during the breaking of the Earth's surface.
Abrupt stopping is usually caused by rupture complexity.
For example, dynamic rupture decelerates quickly when tunneling underneath an orthogonal intersecting fault in the simulations of the 2019 Ridgecrest sequence.
Other rupture complexities that can cause localized high-frequency radiation are fault kinks or sudden changes in along-strike fault geometry, as we observe in the dynamic rupture model of the 1992 Landers earthquake at the JVF--KF and the KF--HVF fault intersections.
Our findings can provide a physical explanation of observations that fault ``misalignment'' \citep{ChuEtAl2021} correlates with enhanced high-frequency seismic radiation, due to a typically higher degree of geometric complexity including more intersecting faults and fault kinks. 

Near-field corner frequency analysis may help to constrain slip distribution and slip segmentation at depth.
Such analysis can also help correlate fault segments with respective subevents in the moment rate release function of large earthquakes.
We find that high-$f_c$ rays can indicate locations of surface rupture. 
Deconvolution of the observed regional ground motions at a station located perpendicular to the rupture direction of the 1992 Landers earthquake shows that surface offsets correlate with the on-fault slip distribution and structural complexity \citep[e.g.][]{KanamoriEtAl1992, Milliner2015}.
The moment rate function of the Landers dynamic rupture model shows that it consists of four sharply separated subevents (see Figure \ref{fig:landers_overview_plot}b, \citealp{KaganHouston2005}; \citealp{ValleeDouet2016}).
Each of these high-$f_c$ rays that we identify can be clearly associated with one of these subevents.
Even the weak moment rate subevent corresponding to the Camp Rock fault rupture is clearly detectable in the $f_c$ distribution. 
Since this rupture on the Camp Rock fault is dynamically triggered in our dynamic rupture scenario and spatially separated from the other slipping parts of the fault system, its associated spectral contribution includes complete nucleation and stopping phases and superimposes the overall source spectrum with a secondary spectrum with a higher $f_c$. 

While our results show that source complexity strongly affects $f_c$ in the near-field, the impact of source complexity on far-field corner frequencies, and thus Brune-type stress drop estimates, may also be larger than typically assumed from simpler rupture models (e.g., \citealp{KanekoShearer2014}; \citealp{WangDay2017}).
\citet{LiuEtAl2023} found that Brune-type corner frequencies of the spectra of the source time functions of complex events correlate best with the corner frequency of that subevent with the highest moment release.
They conclude that the Brune stress drop reflects the stress change of the largest asperity.
Our observed spatial variability of $f_c$ in dynamic rupture simulations paints an even more complex picture, identifying dynamic earthquake characteristics as an important source of ground motion spectra variability.
Recent observations align with our findings: \citet{CalderoniAbercrombie2023} compare stress drops inferred from finite-fault modeling with Brune-type corner frequencies for the M$_w$ 6.0 Amatrice and the M$_w$ 5.9 Visso events in Italy.
They find that high spectral corner frequencies may not be directly associated with high stress drops but rather reflect high-frequency ground motions caused by rupture complexity.
Our results highlight the importance of good azimuthal coverage when inferring Brune-type stress drops from corner frequencies \citep{KanekoShearer2015}. 
Future larger-scale, full-complexity dynamic rupture simulations generating realistic high-resolution ground motion synthetics can further validate dynamic source effects on far-field stress drops.

\citet{Umeda1990} introduced the concept of an ``earthquake bright spot'' as a localized area in the shallow fault region that emits strong high-frequency waves. 
Specifically, the Landers earthquake fault system kink where the JVF branches into the KF has been identified as an earthquake bright spot \citep{YamashitaUmeda1994} which agrees with our dynamic rupture model analysis, where it is a prominent source of a ray of elevated $f_c$.
\citet{YamashitaUmeda1994} propose that the nucleation and arrest of slip on subsidiary faults cause earthquake bright spots.
We find that rake-rotated along-strike surface-breaking rupture fronts that encounter geometric fault complexities can locally cause strong acceleration pulses, which can equally explain the origin of earthquake bright spots without the need for secondary faults.

In addition, tossed-up boulders indicate that co-seismic vertical accelerations exceeded gravity during the 1992 Landers earthquake \citep{YamashitaUmeda1994} and similar high vertical accelerations were recorded or inferred for other large strike-slip earthquakes (e.g., \citealp{Archuleta1982}; \citealp{StrasserBommer2009}; \citealp{Kaiser2017}; \citealp{Hough2020}). 
Numerous vertical acceleration recordings of reverse faulting earthquakes exceeded gravity (e.g., \citealp{BilhamEngland2001}; \citealp{Causse2021}), which is not unexpected as the vertical components are affected stronger by SH-waves. 
We demonstrate how complex source mechanisms of surface-rupturing strike-slip events can cause strong vertical acceleration pulses that may locally exceed gravity.

Equivalent near-field corner frequency analysis of the Mw 6.4 Searles Valley model showcases the capability of $f_c$ variability to track major path effects.
A mountain range deflects a high-frequent wave packet and directs seismic energy in an unexpected direction.
We find evidence of this deflection in observed waveforms, e.g., at station WMF (Figure \ref{fig:RC_e1_fc_map_3x1}a-c), where this effect doubles the shaking duration. 
Such path effects can be relevant for seismic hazard assessment but may be missed in ground motion models that do not include variability in shaking duration. 

Although each presented $f_c$ distribution map is inferred from more than 1,000,000 virtual stations, important aspects of our findings are equally inferrable from lower resolution analysis, e.g., using only 1\% of the data ($\sim$10,000 virtual stations with a spacing of $\sim$5\,km, Figure S4) which is promising for potential real-world applicability of the method. 
For example, the six high-$f_c$ rays pointing away from the fault trace and the sharp $f_c$ increase at the Salton Sea basin in the South and the San Bernardino and Los Angeles basins in the South-West, are still clearly visible in the low-resolution version of the vertical $f_c$ distribution of the Landers model (Figure S4).
Similarly dense seismic sensor locations are becoming feasible. 
For example, the LArge-n Seismic Survey in Oklahoma (LASSO) experiment deployed more than 1800 vertical component nodal seismometers covering a 25\,km by 32\,km region with a station spacing of $\sim$400\,m \citep{DoughertyEtAl2019}.
Distributed Acoustic Sensing (DAS) can provide linear arrays with a sensor spacing of $\sim$10\,m \citep[e.g.,][]{Zhan2019}.

\section*{Conclusions} \label{sec:Conclusions}
In this study, we present a detailed analysis of the spatial variability of an equivalent near-field corner frequency $f_c$ in large strike-slip 3D dynamic rupture simulations.
We discover patterns of highly variably $f_c$ and show that $f_c$ variability is dominantly controlled by source effects.
Rays of locally increased $f_c$ values radiate outwards from the dynamically slipping faults, particularly noticeable in the vertical components.
We validate the variability in the distribution of $f_c$ from the dynamic rupture model with those derived from observed spectra.
We use isochrone analysis to show that the radiation of vertical high-frequencies often results from rake-rotated surface-breaking rupture fronts that decelerate suddenly due to source complexities, such as fault heterogeneities or geometric complexity.
We observe that the P-SV radiation pattern, in combination with the directivity effect, shapes high-$f_c$ rays at a 45$\degree$ angle to the forward rupture propagation direction.
This dynamic source effect can potentially explain observations of high-intensity, impulsive near-field vertical ground motions. 
The analyses of near-field $f_c$ distributions can inform on characteristics of earthquake kinematics and dynamics including rupture directivity, surface rupture, and fault segmentation. 
We find that path effects additionally imprint on the dynamic rupture equivalent corner frequencies of near-field spectra.
For example, we observe a strong deflection of a high-$f_c$ ray along the southern Sierra Nevada mountain range in the Mw 6.4 Searles Valley simulation. 
In conclusion, our findings highlight that the equivalent near-field corner frequency may serve as an insightful ground motion parameter. 
$f_c$ can be inferred from spatially dense, relatively low-frequency ground motion data sets, thereby offering an approach to directly infer the spectral fingerprints of rupture dynamics from near-field ground motions.
Our study has important implications for seismic hazard assessment and offers new avenues for interpreting large array or Distributed Acoustic Sensing data to improve our understanding of the dynamics and ground motions of large earthquakes.

\section*{Data and Resources}\label{sec:DataAndResources}
The described complexity of each dynamic rupture earthquake scenario is best illustrated in animations of the dynamic rupture models of the 1992 Mw 7.3 Landers earthquake (\url{https://www.youtube.com/watch?v=zi19g5Jpp5s}), the 2019 Mw 6.4 Searles Valley foreshock (\url{https://www.youtube.com/watch?v=4b_uhs_rT_g}), and the 2019 Mw 7.1 Ridgecrest mainshock (\url{https://www.youtube.com/watch?v=8yP0rcC7n-g}).
The open-source software package SeisSol is available at \url{https://github.com/SeisSol/SeisSol}.
All input files that are needed to run the Ridgecrest models are available at Zenodo (\url{https://zenodo.org/record/6842773}).
The SeisSol branch that was used to run the Ridgecrest models is also archived (\url{https://zenodo.org/record/7642533}).
The script that calculates the equivalent near-field corner frequencies from SeisSol's free surface output is provided in the repository (\url{https://github.com/SeisSol/SeisSol/blob/master/postprocessing/science/spectral_corner_frequency_from_surface_xdmf.py}).
The script that was used to compute isochrons directly from SeisSol's raw output data is openly available in the SeisSol repository (\url{https://github.com/SeisSol/SeisSol/blob/master/postprocessing/science/compute_isochrones.py}).
The scripts use the external libraries NumPy and SciPy (\citealp{HarrisEtAl2020}; \citealp{VirtanenEtAl2020}).
Details about the TPV5 benchmark problem are provided on the homepage of the SCEC/USGS rupture dynamics code verification community effort (\url{https://strike.scec.org/cvws/tpv5docs.html}).
All seismic data were downloaded through the IRIS Wilber 3 system (https://ds.iris.edu/wilber3/) from the Southern California Seismic Network \citep[CI,][]{SCSN}.
The Python package ObsPy was used to remove the instrument response \citep{Krischeretal2015}.
The electronic supplement contains four additional figures.

\section*{Declaration of Competing Interests}
The authors declare no competing interests.

\section*{Acknowledgments}\label{sec:Acknowledgments}
 
We thank Franti\v{s}ek Gallovi\v{c}, Jean-Paul Ampuero and P. Martin Mai for inspiring discussions; and Stephanie Wollherr and Taufiq Taufiqurrahman for discussing their dynamic rupture models.
This study was supported by the European Union’s Horizon 2020 Research and Innovation Programme (TEAR grant number 852992), Horizon Europe (ChEESE-2P grant number 101093038, DT-GEO grant number 101058129, and Geo-INQUIRE grant number 101058518), the National Aeronautics and Space Administration (80NSSC20K0495), the National Science Foundation (grant No. EAR-2121666) and the Southern California Earthquake Center (SCEC awards 22135, 23121).
We would like to express our gratitude to the Gauss Centre for Supercomputing e.V. (www.gauss-centre.eu) for providing us with computing time on the supercomputer SuperMUC-NG at the Leibniz Supercomputing Centre (www.lrz.de) for our project pr63qo.
Additional resources were provided by the Institute of Geophysics of LMU Munich \citep{Oeser2006}.

\newpage
\bibliographystyle{abbrvnat}
\bibliography{References}

\newpage

nico.schliwa@geophysik.uni-muenchen.de

algabriel@ucsd.edu

\newpage
\section*{List of Figure Captions}\label{sec:FigureCaptions}

\begin{figure}[!htbp] 
\centering
\includegraphics[scale=0.53]{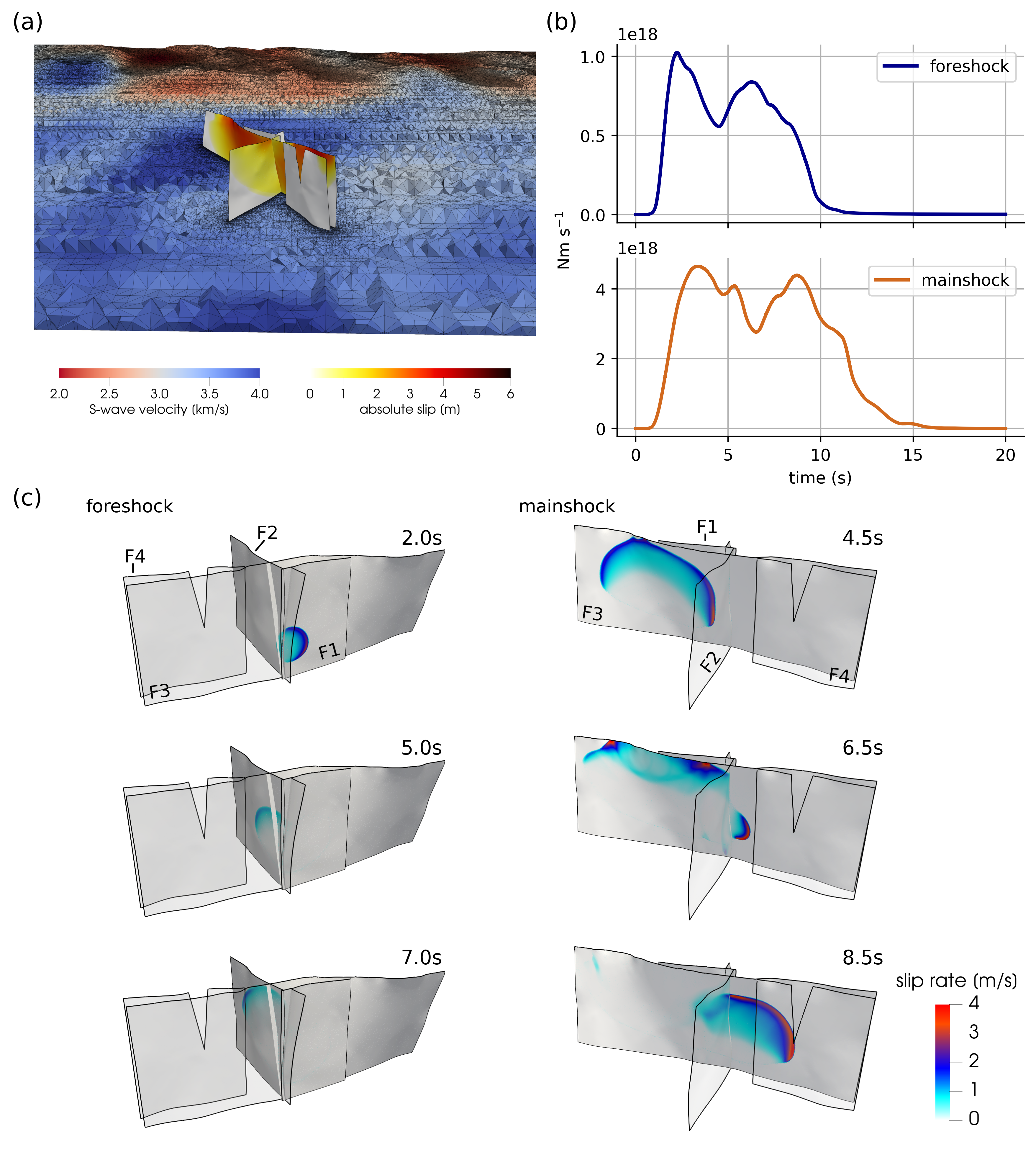}
\caption{Overview of the 2019 Ridgecrest sequence (linked $M_w$ 6.4 Searles Valley foreshock and $M_w$ 7.1 Ridgecrest mainshock) 3D dynamic rupture models adapted from \citet{TaufiqEtAl2023}. (a) Fault geometry with slip distribution after both earthquakes and cross-cut of the unstructured tetrahedral computational mesh colored by the used 3D variable S-wave velocity \citep[CVM-S4.26; ][]{Leeetal2014}. (b) Seismic moment release rate for both, foreshock and mainshock. (c) Slip rate snapshots across the orthogonal fault system at selected rupture times, illustrating dynamic rupture evolution and complexity. The foreshock dynamic rupture scenario is shown on the left side and the mainshock is on the right side.}
\label{fig:ridgecrest_overview_plot}
\end{figure}

\begin{figure}[!htb] 
\centering
\includegraphics[scale=0.38]{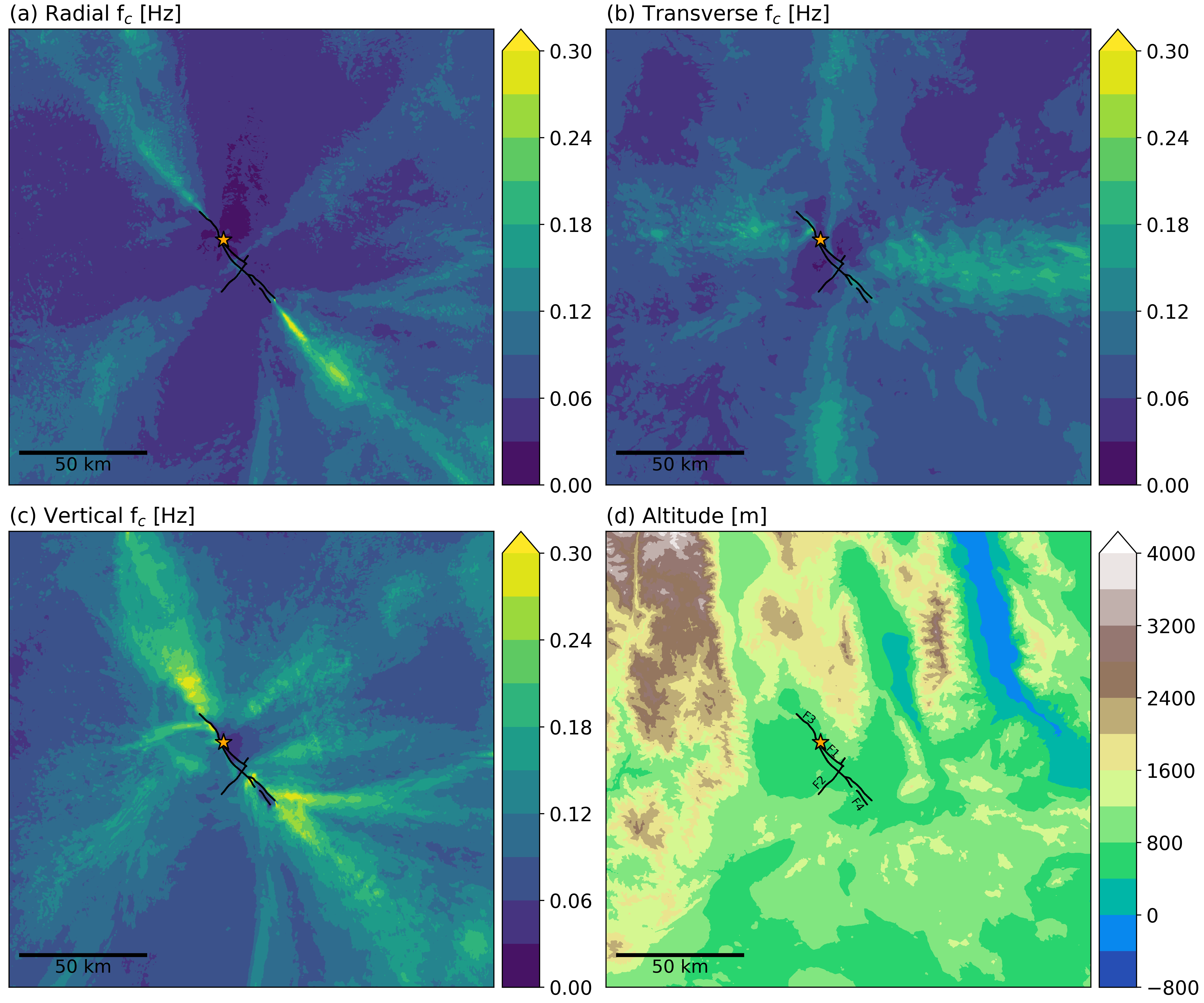}
\caption{Map view of the equivalent near-field corner frequency ($f_c$) distribution of the (a) radial, (b) transverse, and (c) vertical components of synthetic seismograms simulated at $\sim$1,800,000 virtual seismic stations. The synthetic seismograms are generated from the complex 3D dynamic rupture model of the 2019 Ridgecrest mainshock (Figure \ref{fig:ridgecrest_overview_plot}).% \citep[Fig. \ref{fig:ridgecrest_overview_plot}, ][]{TaufiqEtAl2023}. 
(d) Map view of the model's topography. Black lines indicate the numbered fault traces and the star marks the epicenter.}
\label{fig:RC_e2_bw_fc_2x2}
\end{figure}

\begin{figure}[!htb] 
\centering
\includegraphics[scale=0.38]{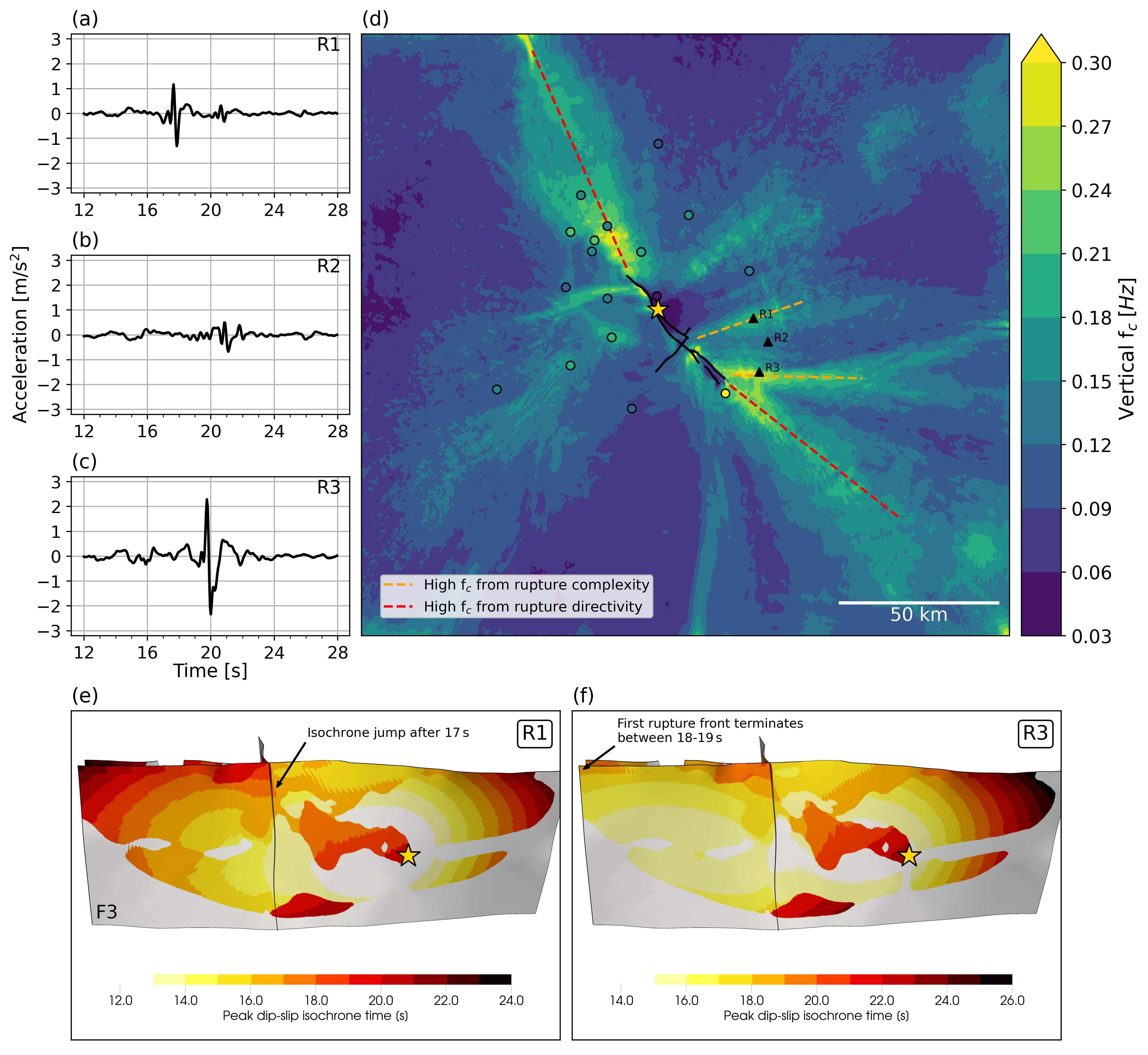}
\caption{(a,b,c) Synthetic vertical ground accelerations at three selected stations. (d) Map view of the equivalent near-field corner frequency ($f_c$) distribution of the vertical components of synthetic seismograms recorded at $\sim$1,800,000 virtual seismic stations. The synthetic seismograms are generated from a complex dynamic rupture model of the 2019 Ridgecrest mainshock (Figure \ref{fig:ridgecrest_overview_plot}). Black lines indicate the fault traces, the star marks the epicenter, colored dots show $f_c$ values of recorded ground motion spectra, and triangles show the virtual station locations of the analyzed accelerograms. Orange and red lines mark different high-$f_c$ features. (e,f) Peak dip-slip isochrones of stations R1 and R3.}
\label{fig:RC_e2_fc_map_3x1}
\end{figure}

\begin{figure}[!htb] 
\centering
\includegraphics[scale=0.4]{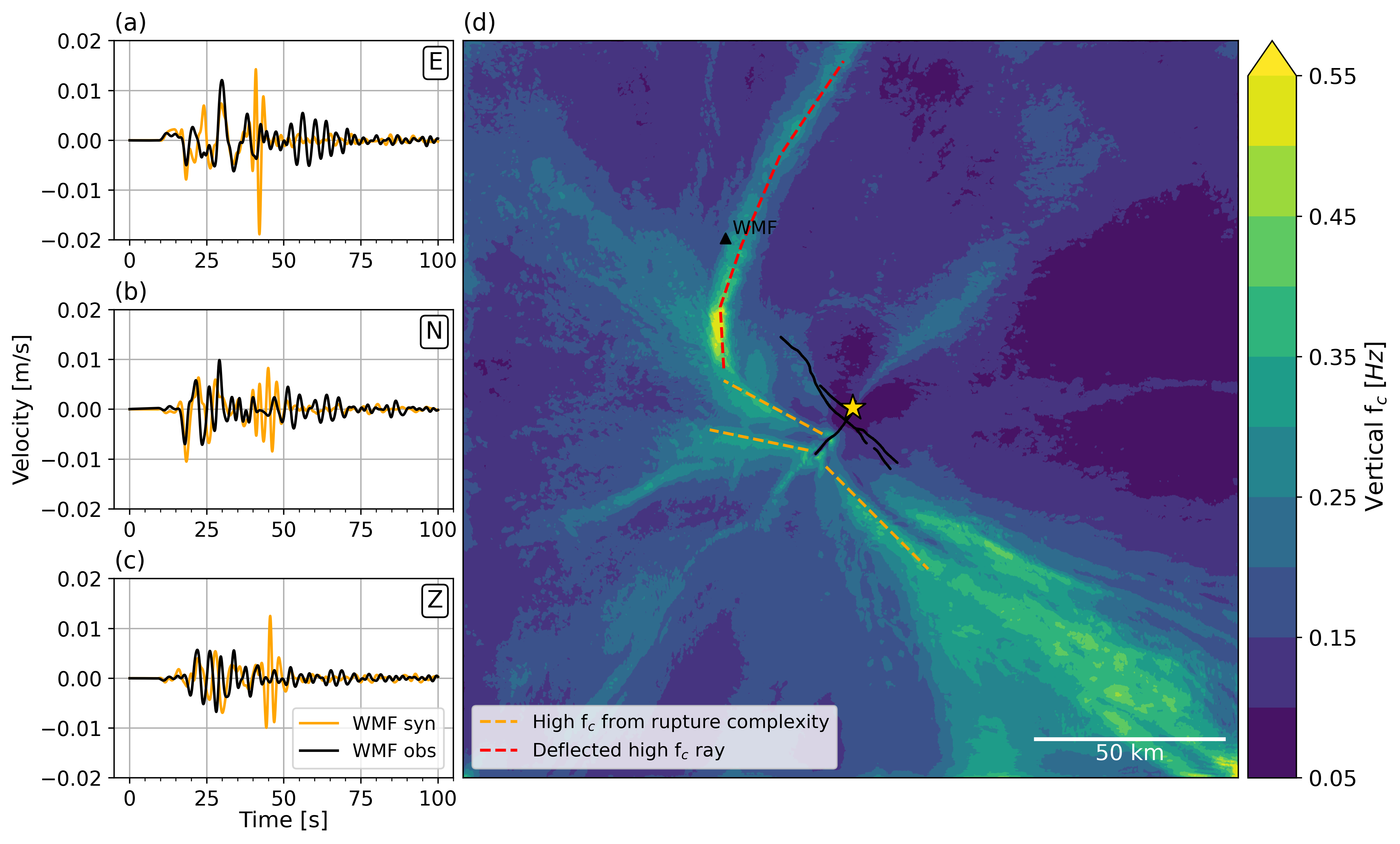}
\caption{Synthetic and observed velocity seismograms of the (a) East, (b) North, and (c) Up components of the Searles Valley foreshock at station WMF. In difference to \cite{TaufiqEtAl2023}, we here show seismograms not normalized and including higher frequencies, lowpass-filtered to 0.5\,Hz to highlight the match of the first wave packet. (d) Map view of the equivalent near-field corner frequency ($f_c$) distribution of the vertical components of synthetic seismograms simulated at $\sim$1,800,000 virtual seismic stations (without picking a body-wave window). The seismograms are generated from the complex 3D dynamic rupture model of the 2019 Searles Valley forehock (Figure \ref{fig:ridgecrest_overview_plot}). Black lines indicate the fault traces, the star marks the epicenter, and the triangle shows the location of the station WMF.}
\label{fig:RC_e1_fc_map_3x1}
\end{figure}

\begin{figure}[!htb] 
\centering
\includegraphics[scale=0.53]{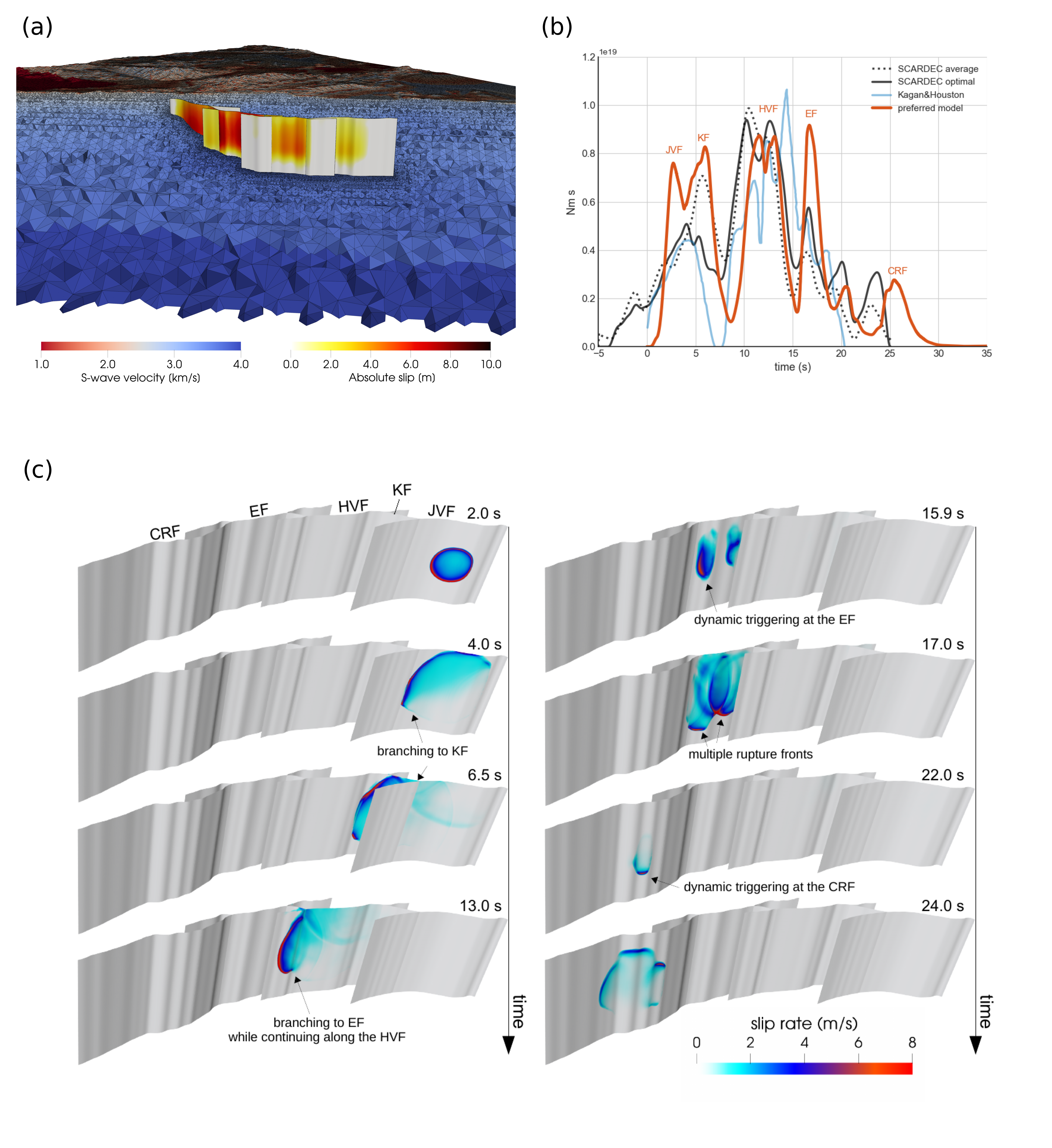}
\caption{Overview of the 1992 Landers earthquake 3D dynamic rupture model adapted from \citet{Wollherretal2019}. (a) Fault geometry with accumulated slip distribution and cross-cut through the unstructured tetrahedral computational mesh colored by the used 3D variable S-wave velocity \citep{Shawetal2015}. (b) Seismic moment release rate. The Landers dynamic rupture model (preferred model, orange) is compared to the optimal and average moment rate release of the SCARDEC database (in black and dotted light gray, \citealp{ValleeDouet2016}) and the inferred moment rate based on the surface slip (in light blue, \citealp{KaganHouston2005}). CRF = Camp Rock Fault; EF = Emerson Fault; HVF = Homestead Valley Fault; KF = Kickapoo Fault; JVF = Johnson Valley Fault. (c) Slip rate snapshots across the fault system at selected rupture times illustrating dynamic rupture evolution and complexity. Rupture cascades across fault segments through direct branching and dynamic triggering.}
\label{fig:landers_overview_plot}
\end{figure}

\begin{figure}[!htb] 
\centering
\includegraphics[scale=0.38]{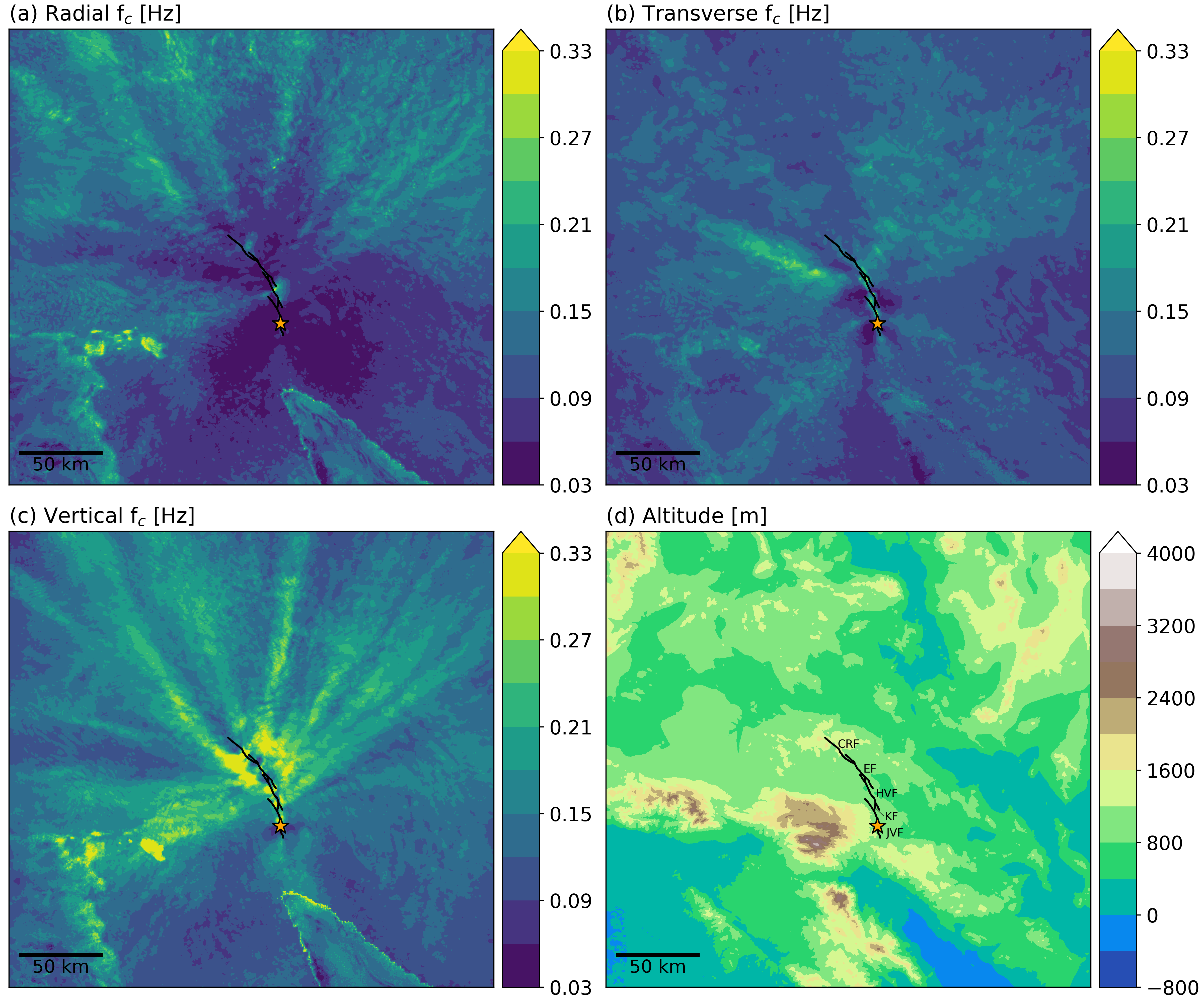}
\caption{Map view of the equivalent near-field corner frequency ($f_c$) distribution of the (a) radial, (b) transverse, and (c) vertical components of synthetic seismograms simulated at $\sim$1,000,000 virtual seismic stations. The seismograms are generated from the complex 3D dynamic rupture model of the 1992 Landers earthquake (Figure \ref{fig:landers_overview_plot}). We clip the color map at sedimentary basins and close to the fault, where static displacement and an inaccurate component separation due to finite-fault effects distort the corner frequency determination. We omit these regions in our interpretation. (d) Map-view of the model's topography. Black lines indicate the fault traces and the star marks the epicenter (JVF = Johnson Valley fault, KF = Kickapoo fault, HVF = Homestead Valley fault, EF = Emmerson fault, CRF = Camp Rock fault).}
\label{fig:Landers_fc_2x2}
\end{figure}

\begin{figure}[!htb] 
\centering
\includegraphics[scale=0.4]{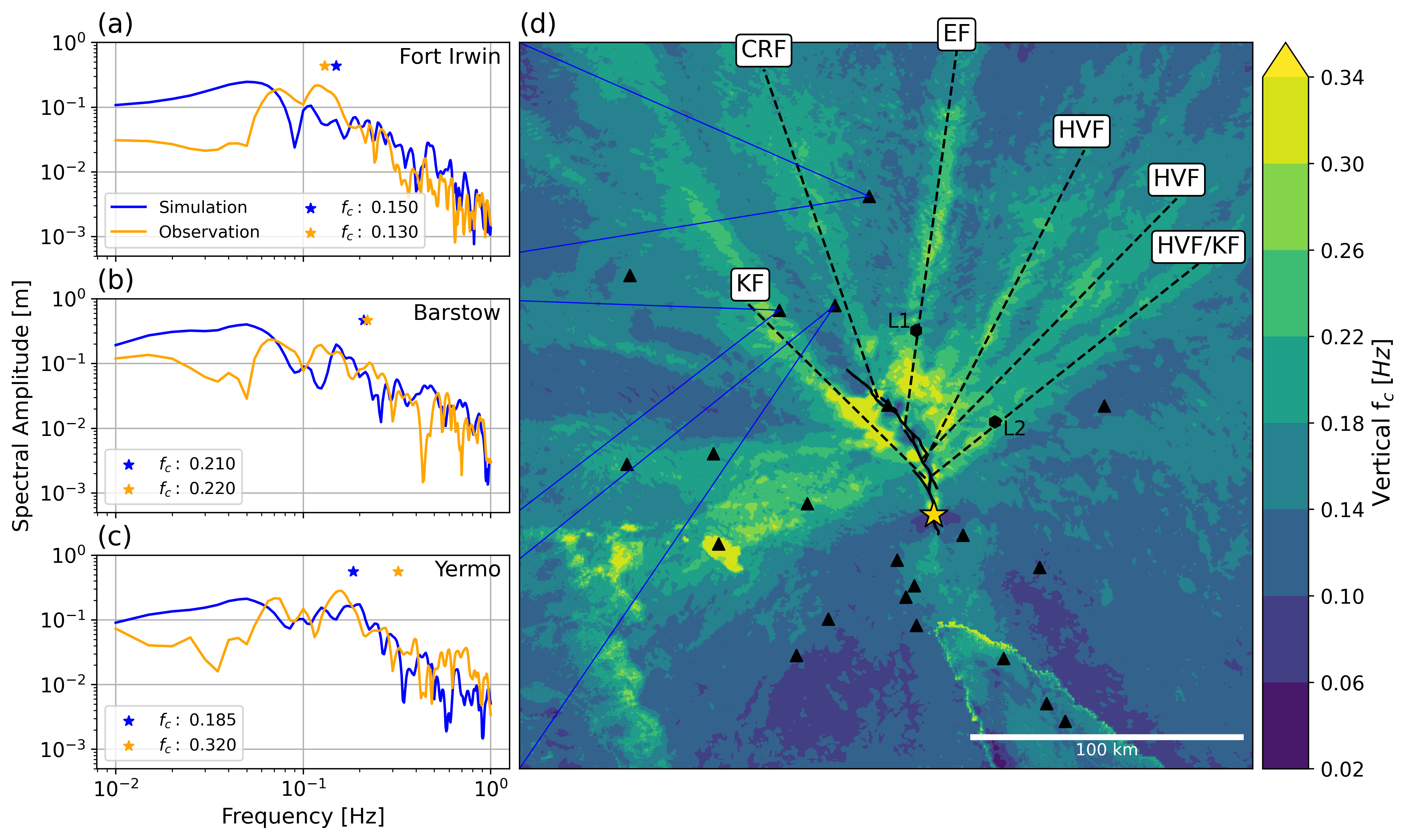}
\caption{Observed spectra and corresponding $f_c$ of the 1992 Landers earthquake compared to synthetic counterparts at three selected stations: (a) Fort Irwin, (b) Barstow, (c) Yermo. The spectra are not normalized but reflect absolute values.
%(d) Map view of the equivalent near-field corner frequency ($f_c$) distribution of the vertical components of synthetic seismograms simulated at $\sim$1,000,000 virtual seismic stations. The seismograms are generated from a complex dynamic rupture model of the 1992 Mw 7.3 Landers earthquake \citep{Wollherretal2019}. 
(d) Same as Figure \ref{fig:Landers_fc_2x2}c. Solid black lines indicate the fault traces, the star marks the epicenter, triangles are real station locations, and hexagons show two virtual stations that are analyzed in Figure \ref{fig:Landers_isochrones}. Dashed lines highlight rays of high $f_c$ and the text windows show the fault names where the rays originate (KF = Kickapoo fault, HVF = Homestead Valley fault, EF = Emmerson fault, CRF = Camp Rock fault). We clip the color map at sedimentary basins and close to the fault, where static displacement and an inaccurate component separation due to finite-fault effects distort the corner frequency determination. We omit these regions in our interpretation.}
\label{fig:Landers_fc_map_3x1}
\end{figure}

\begin{itemize}
  \item Figure \ref{fig:tpv5_fc_4x2.png}: Equivalent near-field corner frequency ($f_c$) distribution of the (a) radial, (b) transverse, and (c) vertical components of synthetic seismograms recorded at $\sim$900,000 virtual seismic stations in map view. The seismograms are generated in a bilateral strike-slip 3D dynamic rupture model including an asperity and a barrier embedded in a homogeneous elastic half-space (the TPV5 SCEC/USGS community benchmark \citep{HarrisEtAl2009}). Black lines indicate the fault trace, the star marks the hypocenter, and white triangles are stations that are analyzed in subplots e-h. Orange lines mark different high-$f_c$ features. (d) Side-view of the fault plane with rupture front contours in 0.5\,s intervals. (e) Isochrone contours of station T1 in 0.5\,s intervals. (f) Transverse ground accelerations at station T1. Comparison to isochrones allows associating pronounced high-amplitude signals with different stages of 3D dynamic rupture propagation. Acc. = Acceleration; dec. = Deceleration; asp. = Asperity. (g)  Peak dip-slip isochrone contours of station T2 in 0.5\,s intervals. (h) Vertical ground accelerations at station T2.
  
  \item Figure \ref{fig:rake_rotation_figure}: (a) Rake at t = 6\,s of the TPV5 dynamic rupture model \citep{HarrisEtAl2009}. (b) Average rake of the 2019 $M_w$ 7.1 Ridgecrest mainshock dynamic rupture model \citep{TaufiqEtAl2023}. (c) Average rake of the 1992 $M_w$ 7.3 Landers dynamic rupture model \citep{Wollherretal2019}.
  
  \item Figure \ref{fig:ridgecrest_overview_plot}: Overview of the 2019 Ridgecrest sequence (linked $M_w$ 6.4 Searles Valley foreshock and $M_w$ 7.1 Ridgecrest mainshock) 3D dynamic rupture models adapted from \citet{TaufiqEtAl2023}. (a) Fault geometry with slip distribution after both earthquakes and cross-cut of the unstructured tetrahedral computational mesh colored by the used 3D variable S-wave velocity \citep[CVM-S4.26; ][]{Leeetal2014}. (b) Seismic moment release rate for both, foreshock and mainshock. (c) Slip rate snapshots across the orthogonal fault system at selected rupture times, illustrating dynamic rupture evolution and complexity. The foreshock dynamic rupture scenario is shown on the left side and the mainshock is on the right side.
  
  \item Figure \ref{fig:RC_e2_bw_fc_2x2}: Map view of the equivalent near-field corner frequency ($f_c$) distribution of the (a) radial, (b) transverse, and (c) vertical components of synthetic seismograms simulated at $\sim$1,800,000 virtual seismic stations. The synthetic seismograms are generated from the complex 3D dynamic rupture model of the 2019 Ridgecrest mainshock (Figure \ref{fig:ridgecrest_overview_plot}). (d) Map view of the model's topography. Black lines indicate the numbered fault traces and the star marks the epicenter.
  
  \item Figure \ref{fig:RC_e2_fc_map_3x1}: (a,b,c) Synthetic vertical ground accelerations at three selected stations. (d) Map view of the equivalent near-field corner frequency ($f_c$) distribution of the vertical components of synthetic seismograms recorded at $\sim$1,800,000 virtual seismic stations. The synthetic seismograms are generated from a complex dynamic rupture model of the 2019 Ridgecrest mainshock (Figure \ref{fig:ridgecrest_overview_plot}). Black lines indicate the fault traces, the star marks the epicenter, colored dots show $f_c$ values of recorded ground motion spectra, and triangles show the virtual station locations of the analyzed accelerograms. Orange and red lines mark different high-$f_c$ features. (e,f) Peak dip-slip isochrones of stations R1 and R3.
  
  \item Figure \ref{fig:RC_e1_fc_map_3x1}: Synthetic and observed velocity seismograms of the (a) East, (b) North, and (c) Up components of the Searles Valley foreshock at station WMF. In difference to \cite{TaufiqEtAl2023}, we here show seismograms not normalized and including higher frequencies, lowpass-filtered to 0.5\,Hz to highlight the match of the first wave packet. (d) Map view of the equivalent near-field corner frequency ($f_c$) distribution of the vertical components of synthetic seismograms simulated at $\sim$1,800,000 virtual seismic stations (without picking a body-wave window). The seismograms are generated from the complex 3D dynamic rupture model of the 2019 Searles Valley forehock (Figure \ref{fig:ridgecrest_overview_plot}). Black lines indicate the fault traces, the star marks the epicenter, and the triangle shows the location of the station WMF.
  
  \item Figure \ref{fig:landers_overview_plot}: Overview of the 1992 Landers earthquake 3D dynamic rupture model adapted from \citet{Wollherretal2019}. (a) Fault geometry with accumulated slip distribution and cross-cut through the unstructured tetrahedral computational mesh colored by the used 3D variable S-wave velocity \citep{Shawetal2015}. (b) Seismic moment release rate. The Landers dynamic rupture model (preferred model, orange) is compared to the optimal and average moment rate release of the SCARDEC database (in black and dotted light gray, \citealp{ValleeDouet2016}) and the inferred moment rate based on the surface slip (in light blue, \citealp{KaganHouston2005}). CRF = Camp Rock Fault; EF = Emerson Fault; HVF = Homestead Valley Fault; KF = Kickapoo Fault; JVF = Johnson Valley Fault. (c) Slip rate snapshots across the fault system at selected rupture times illustrating dynamic rupture evolution and complexity. Rupture cascades across fault segments through direct branching and dynamic triggering.
  
  \item Figure \ref{fig:Landers_fc_2x2}: Map view of the equivalent near-field corner frequency ($f_c$) distribution of the (a) radial, (b) transverse, and (c) vertical components of synthetic seismograms simulated at $\sim$1,000,000 virtual seismic stations. The seismograms are generated from the complex 3D dynamic rupture model of the 1992 Landers earthquake (Figure \ref{fig:landers_overview_plot}). We clip the color map at sedimentary basins and close to the fault, where static displacement and an inaccurate component separation due to finite-fault effects distort the corner frequency determination. We omit these regions in our interpretation. (d) Map-view of the model's topography. Black lines indicate the fault traces and the star marks the epicenter (JVF = Johnson Valley fault, KF = Kickapoo fault, HVF = Homestead Valley fault, EF = Emmerson fault, CRF = Camp Rock fault).
  
  \item Figure \ref{fig:Landers_fc_map_3x1}: Observed spectra and corresponding $f_c$ of the 1992 Landers earthquake compared to synthetic counterparts at three selected stations: (a) Fort Irwin, (b) Barstow, (c) Yermo. The spectra are not normalized but reflect absolute values. (d) Same as Figure \ref{fig:Landers_fc_2x2}c. Solid black lines indicate the fault traces, the star marks the epicenter, triangles are real station locations, and hexagons show two virtual stations that are analyzed in Figure \ref{fig:Landers_isochrones}. Dashed lines highlight rays of high $f_c$ and the text windows show the fault names where the rays originate (KF = Kickapoo fault, HVF = Homestead Valley fault, EF = Emmerson fault, CRF = Camp Rock fault). We clip the color map at sedimentary basins and close to the fault, where static displacement and an inaccurate component separation due to finite-fault effects distort the corner frequency determination. We omit these regions in our interpretation.
  
  \item Figure \ref{fig:Landers_isochrones}: (a,c,e) Synthetic vertical accelerograms at three selected stations: Yermo, L1, and L2 (Figure \ref{fig:Landers_fc_map_3x1}d). (b,d,f) Peak dip-slip isochrones of the respective stations. The star marks the hypocenter (JVF = Johnson Valley fault, KF = Kickapoo fault, HVF = Homestead Valley fault, EF = Emmerson fault, CRF = Camp Rock fault).
\end{itemize}

\newpage
\section*{Figures}\label{sec:Figures}

\begin{figure}[!htb] 
\centering
\includegraphics[scale=0.4]{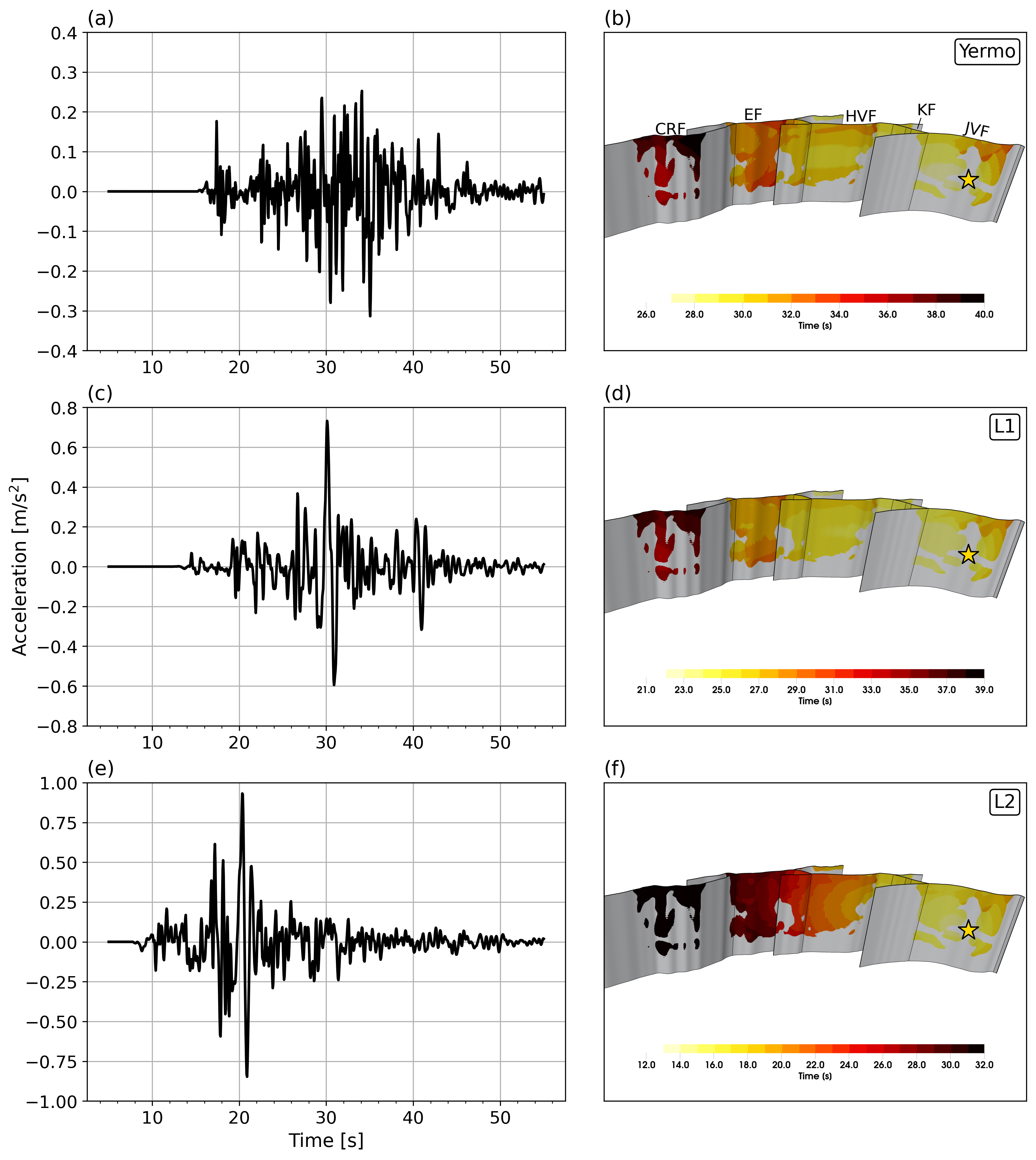}
\caption{(a,c,e) Synthetic vertical accelerograms at three selected stations: Yermo, L1, and L2 (Figure \ref{fig:Landers_fc_map_3x1}d). (b,d,f) Peak dip-slip isochrones of the respective stations. The star marks the hypocenter (JVF = Johnson Valley fault, KF = Kickapoo fault, HVF = Homestead Valley fault, EF = Emmerson fault, CRF = Camp Rock fault).}
\label{fig:Landers_isochrones}
\end{figure}

\renewcommand{\thefigure}{A\arabic{figure}}
\setcounter{figure}{0}
\renewcommand{\thetable}{A\arabic{table}}
\setcounter{table}{0}

\newpage
\section*{Appendices}\label{sec:Appendices}
\section*{3D dynamic rupture models }\label{sec:Models}
The TPV5 USGS/SCEC dynamic rupture community benchmark \citep{HarrisEtAl2009} describes a bilateral strike-slip earthquake in 3D dynamically propagating across a rectangular fault that intersects the free surface in an elastic half-space.
The rupture is artificially nucleated in the middle of the fault and then spreads spontaneously in each direction.
To the left/right of its center, the model setup includes two fault asperities/barriers with elevated or reduced initial shear stresses, which accelerate or decelerate dynamic rupture, respectively.
The rupture is forced to stop abruptly at the fault edges.
The dynamic model parameters are summarized in Table \ref{Tab:tpv5_parameters}.

\citet{Wollherretal2019} developed a 3D dynamic rupture model of the 1992 $M_w$ 7.3 Landers earthquake that includes geometric fault complexity and matches a broad range of regional and local observations, including fault slip, seismic moment release, and ground motions (Figure \ref{fig:landers_overview_plot}).
The dynamic rupture earthquake model uses a linear slip-weakening friction law, high-resolution topography, 3D velocity structure (CVM-H; \citealp{Shawetal2015}), viscoelastic attenuation, and off-fault (visco-)plasticity.
The fault system of the Landers dynamic rupture model consists of five vertical subfaults that extend to a depth of 15\,km, adapted from fault traces from photometric images  \citep{Flemingetal1998}.
The fault system exhibits a strike rotation of approximately 30$\degree$, striking towards the north in the southern part and towards the northwest in the northern part.
\citet{Wollherretal2019} find that assuming a constant maximum principal stress direction inhibits rupture propagation across the fault system.
In their setup, the maximum principal stress axis varies smoothly, which is consistent with the regional stress estimates.
Dynamic rupture is initiated by gradually reducing the static friction coefficient to its dynamic value within a circular nucleation patch of radius 1.5\,km \citep{Bizzarri2010}. 
Table \ref{Tab:Landers_parameters} provides an overview of all model parameters.

\citet{TaufiqEtAl2023} modeled linked foreshock-mainshock dynamic rupture scenarios of the 2019 Ridgecrest sequence, the $M_w$ 6.4 Searles Valley foreshock, and the $M_w$ 7.1 Ridgecrest mainshock.
Their dynamic rupture scenario assembles earthquake physics with high-quality strong-motion and teleseismic, field mapping, high-rate GNSS, and space geodetic foreshock and mainshock datasets of Californias's largest earthquakes since 20 years.
The initial 3D stress state is inferred from tectonic background loading \citep{YangandHauksson2013}, which is additionally modulated by long-term Coulomb failure stress changes ($\Delta$CFS) caused by previous major earthquakes in the Eastern California Shear Zone \citep{VerdecchiaandCarena2016}. The prestress of the mainshock dynamic rupture scenario includes stress changes induced by the foreshock.
While a realistic dynamic rupture scenario of the Ridgecrest mainshock needs to account for the stress changes due to the Searles Valley foreshock, the foreshock does not dynamically trigger the mainshock.
The models include viscoelastic attenuation, off-fault plasticity, and a non-vertical quasi-orthogonal 3D fault system with four fault segments (Figure \ref{fig:ridgecrest_overview_plot}).
\citet{TaufiqEtAl2023} construct the fault system geometry by integrating geological field mapping, geodetic InSAR data, relocated seismicity, and selected focal mechanisms \citep{CarenaandSuppe2002}.
The fault system is embedded into a 3D velocity model of southern California (CVM-S4.26; \citealp{Leeetal2014}; \citealp{Smalletal2017}) and intersects high-resolution topography.
Table \ref{tab:Ridgecrest_fault_friction} summarizes the frictional fault properties.
Both ruptures are nucleated by imposing shear stress perturbations in spherical nucleation areas with radii of 3.5\,km around their respective hypocenters. 

All dynamic rupture simulations use the open-source software package SeisSol (see \hyperref[sec:DataAndResources]{Data and Resources}) to solve the problem of spontaneous frictional failure on prescribed faults and non-linearly coupled seismic wave propagation. SeisSol uses the Arbitrary high-order accurate DERivative Discontinuous Galerkin (ADER-DG) method and employs fully adaptive, unstructured tetrahedral meshes \citep{DumbserKaeser2006, DelaPuenteetal2009, Peltiesetal2012}. Tetrahedral elements enable meshing flexibility and are crucial to incorporate complex and intersecting geometries such as those inherent to the Ridgecrest and Landers fault systems. SeisSol is verified in a variety of dynamic rupture benchmark problems (\citealp{Peltiesetal2014}; \citealp{Harrisetal2018}) and is optimized to efficiently exploit high-performance computing infrastructures (e.g., \citealp{Breueretal2014}; \citealp{Uphoffetal2017}, \citealp{KrenzEtAl2021}). SeisSol supports high computational efficiency when incorporating elastic, anisotropic, viscoelastic, viscoplastic, and poroelastic rheologies (\citealp{UphoffBader2016}; \citealp{Wollherretal2018}; \citealp{WolfEtAl2020}; \citealp{WolfEtAl2022}). The Landers and Ridgecrest dynamic rupture earthquake scenarios account for off-fault (visco-)plasticity and viscoelastic attenuation.
Within our here-considered model domains, the seismic wavefield is resolved up to at least 1\,Hz in the TPV5 and Landers models and up to 2\,Hz in the Ridgecrest dynamic rupture simulations.
The TPV5 dynamic rupture model requires 2k CPU hours on the supercomputer Supermuc-NG, the Landers earthquake scenarios required approximately 100k CPU hours on Supermuc Phase-2, and the linked simulation of the Ridgecrest sequence requires 243k CPU hours on Supermuc-NG.

\begin{table}[!htbp]
\centering
\caption{Summary of the TPV5 dynamic rupture model parameters \citep[\protect\url{https://strike.scec.org/cvws/tpv5docs.html};][]{HarrisEtAl2009}.}
\begin{tabular}{ l l l }
Symbol & Parameter & Value\\
\hline
$\mu_s$ & static friction & 0.677 \\
$\mu_d$ & dynamic friction & 0.525 \\
$D_c$ & critical slip-weakening distance & 0.4\,m \\
cohesion & frictional cohesion &  0.0\,MPa \\
s$_{yy}$ & stress & 120\,MPa \\ 
s$_{xx}$, s$_{zz}$, s$_{yz}$, s$_{xz}$ & stress & 0\,MPa \\
s$_{xy}$ & stress outside the nucleation zone & 70\,MPa \\
& stress inside the nucleation zone & 81.6\,MPa \\
& stress inside the barrier & 62\,MPa \\
& stress inside the asperity & 78\,MPa \\
\hline
\end{tabular}
\label{Tab:tpv5_parameters}
\end{table}

\begin{table}[!htbp]
\centering
\caption{Summary of the Landers dynamic rupture model parameters (adapted from Table 1 in \citealp{Wollherretal2019}).}
\begin{tabular}{ l l l }
symbol & parameter & value with units \\
\hline
$\mu_s$ & static friction & 0.55 (0.44 at the EF and CRF)\\
$\mu_d$ & dynamic friction & 0.22 \\
$D_c$ & critical slip distance &0.62~m \\
$c$ & bulk cohesion & depth-dependent, good quality \\
&&  rock model of \cite{RotenEtAl2017}, \\
&&  ranging between 2.5--50.0~MPa \\
$\phi$ & friction angle & 0.55 \\
$v_s, v_p$ & shear and p-wave velocity & 3D CVM-H \cite{Shawetal2015} \\
$\rho$ & density & 3D CVM-H \cite{Shawetal2015}\\
$Q_s, Q_p$ & viscoelastic damping parameters & 50 $v_s$, 2 $Q_s$ \\
$r$ & nucleation patch radius & 1.5~km \\
& forced nucleation time & 0.6~s \\
$R$ & relative pre-stress ratio & 0.65 \\
$\sigma_2$ & principal vertical stress & (2700 -- 1000)~kg/m 9.8 abs(depth~m) \\
$\sigma_1, \sigma_3$ & principal horizontal stresses & amplitudes determined by $R$ and eq. (2) and (3) \\
& & in \cite{Wollherretal2019}\\
$dx$ & smallest element edge & 200~m \\
$p$ & polynomial order of accuracy & 4 \\
\hline
\end{tabular}
\label{Tab:Landers_parameters}
\end{table}

\begin{table*}
\begin{center}
\caption{Rate-and-state frictional fault properties of the Ridgecrest sequence models (adapted from Table S2 in \citealp{TaufiqEtAl2023}).}
\label{tab:Ridgecrest_fault_friction}
\begin{tabular}{ccc}
 \hline
  \textbf{Parameter} & \textbf{Symbol} & \textbf{Value} \\
  \hline
 Direct-effect parameter & \textit{a} & 0.01-0.02 \\ 
  %\hline
 Evolution-effect parameter & \textit{b} & 0.014 \\ 
  %\hline
 Reference slip rate & $V_\mathrm{0}$ & $10^{-6}$ m/s \\ 
  %\hline
 Steady-state low-velocity friction coefficient at the slip rate $V_\mathrm{0}$ & $f_\mathrm{0}$ & 0.6 \\ 
  %\hline
 Characteristic slip distance of the state evolution & $L$ & 0.2 \\ 
  %\hline
 Full weakened friction coefficient & $f_\mathrm{w}$ & 0.1 \\ 
  %\hline
 Initial slip rate & $V_\mathrm{ini}$ & $10^{-16}$ m/s \\ 
  %\hline
 Weakened slip rate & $V_\mathrm{w}$ & 0.1 m/s \\ 
 \hline
\end{tabular}
\end{center}
\end{table*}

\newpage

\title{\Large \textbf{Supplemental Material} \\ \LARGE Equivalent near-field corner frequency analysis of 3D dynamic rupture simulations reveals source complexity}
\author{Nico Schliwa and Alice-Agnes Gabriel}
\maketitle

\newpage

\renewcommand{\thefigure}{S\arabic{figure}}
\setcounter{figure}{0}

\begin{figure}[!htbp] 
\centering
\includegraphics[scale=0.4]{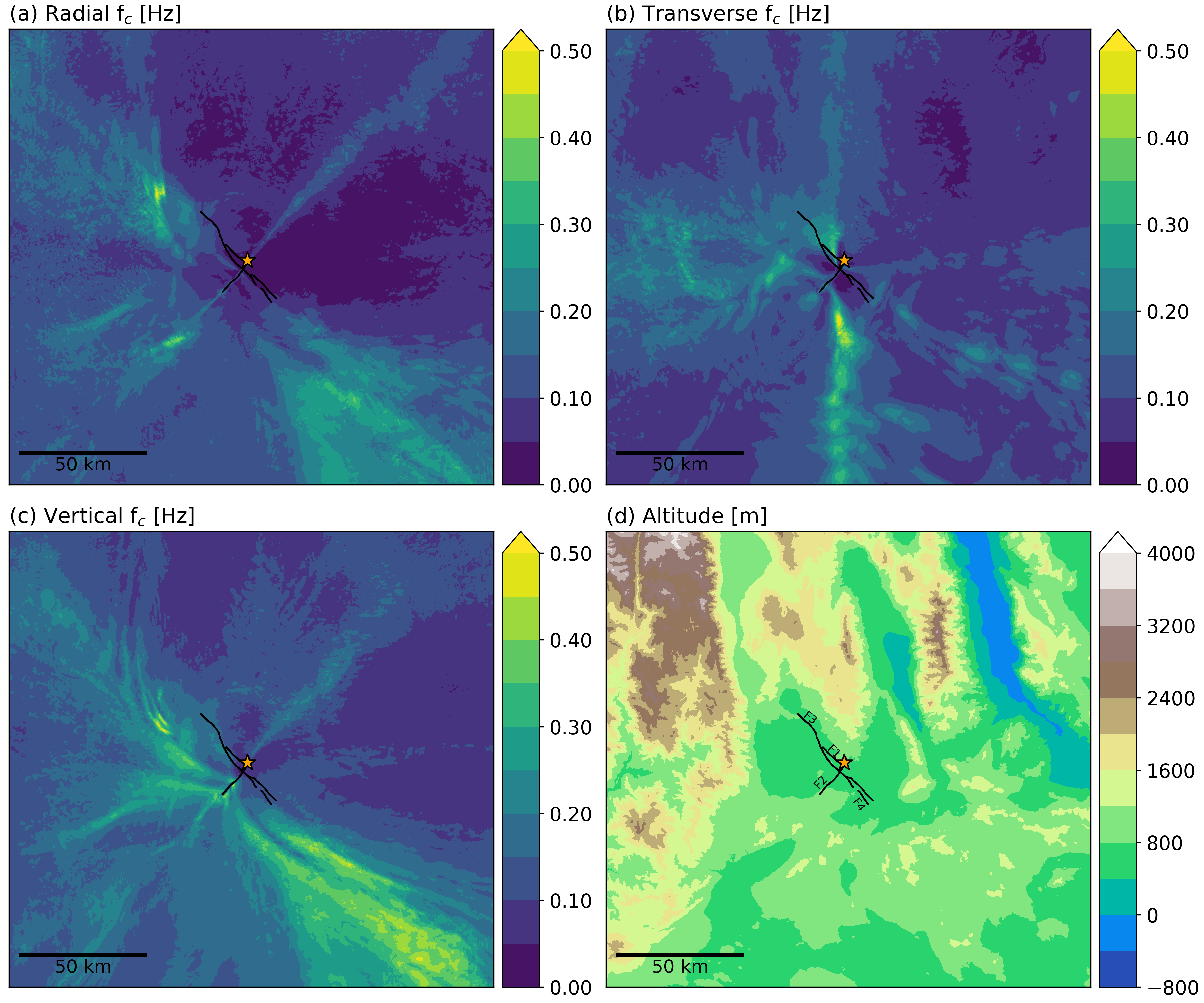}
\caption{Map view of the equivalent near-field corner frequency ($f_c$) distribution of the (a) radial, (b) transverse, and (c) vertical components of synthetic seismograms simulated at $\sim$1,800,000 virtual seismic stations (with picking a body-wave window). The synthetic seismograms are generated from the complex 3D dynamic rupture model of the 2019 Searles Valley foreshock (Figure \ref{fig:ridgecrest_overview_plot}).
(d) Map view of the model's topography. Black lines indicate the numbered fault traces and the star marks the epicenter.}
\label{fig:RC_e1_bw_fc_2x2}
\end{figure}

\begin{figure}[!htbp] 
\centering
\includegraphics[scale=0.4]{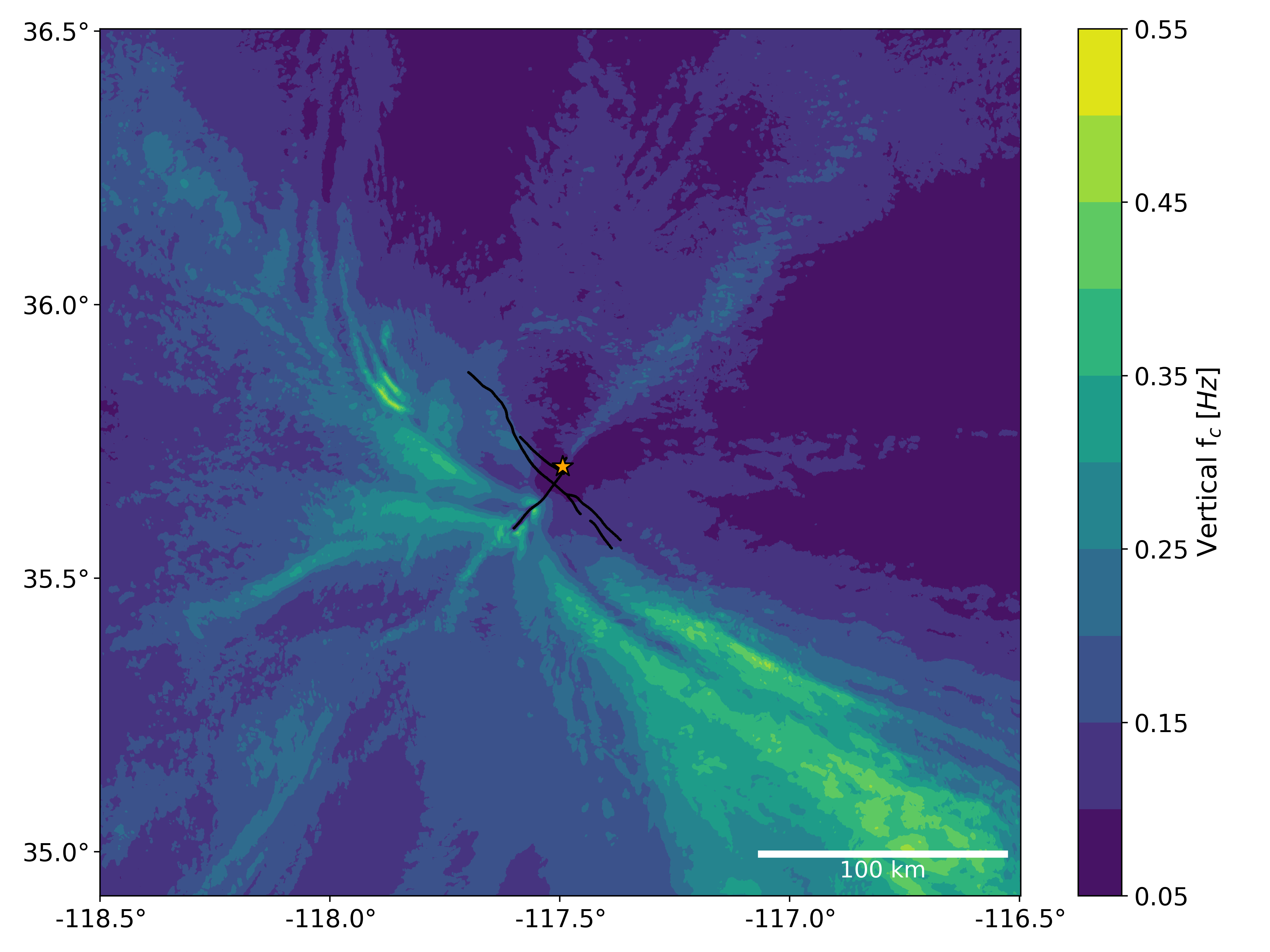}
\caption{Map view of the equivalent near-field corner frequency ($f_c$) distribution of the vertical components of synthetic seismograms simulated at $\sim$1,800,000 virtual seismic stations (with picking a body-wave window). The seismograms are generated from the complex 3D dynamic rupture model of the 2019 Searles Valley forehock (Figure \ref{fig:ridgecrest_overview_plot}). Black lines indicate the fault traces and the star marks the epicenter.}
\label{fig:RC_e1_fc_map}
\end{figure}

\begin{figure}[!htbp] 
\centering
\includegraphics[scale=0.5]{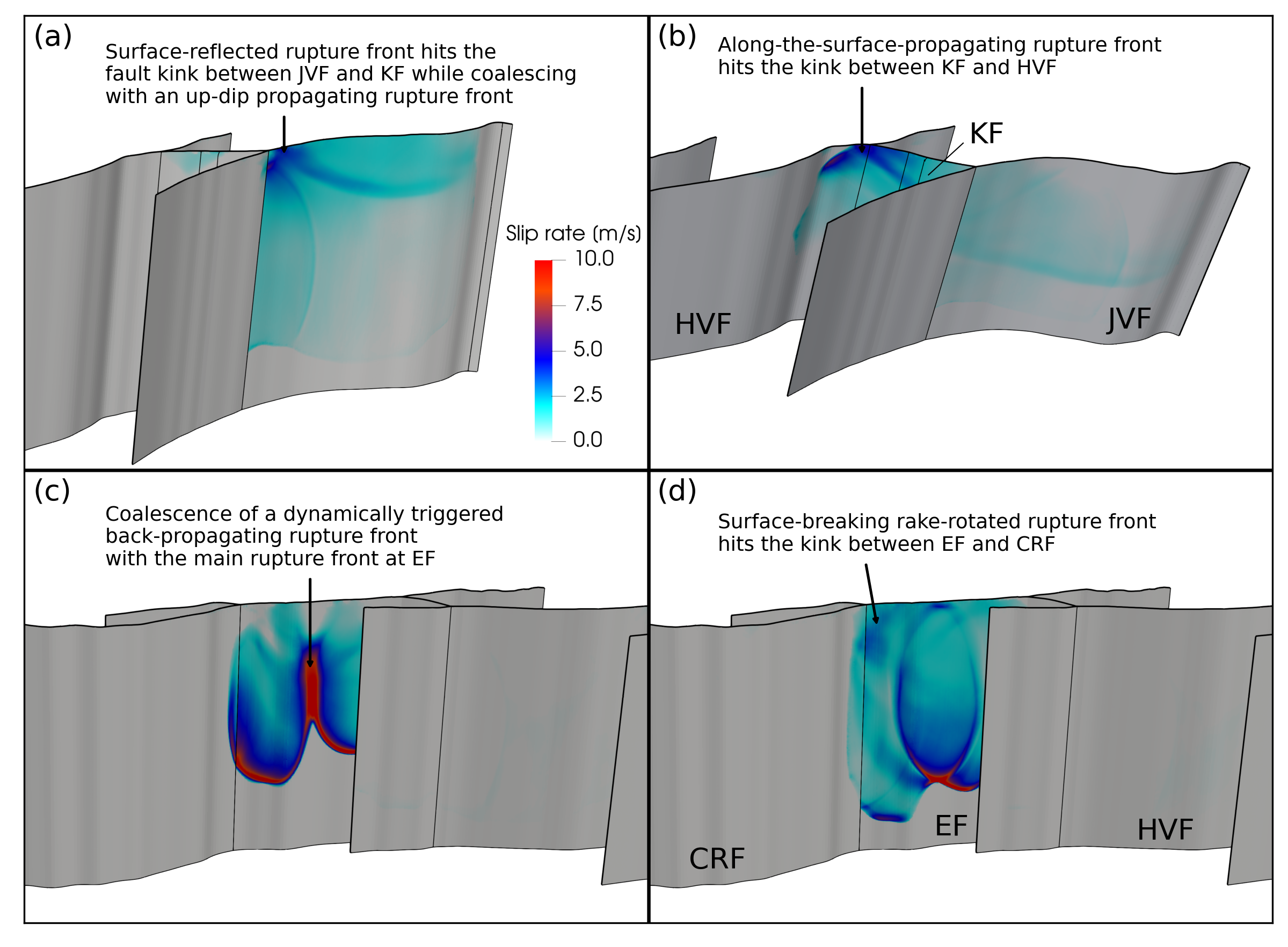}
\caption{Slip rate snapshots of critical rupture phases of the 1992 Landers earthquake 3D dynamic rupture model (Figure \ref{fig:landers_overview_plot}).}
\label{fig:landers_sliprates_supp1}
\end{figure}

\begin{figure}[!htbp] 
\centering
\includegraphics[scale=0.4]{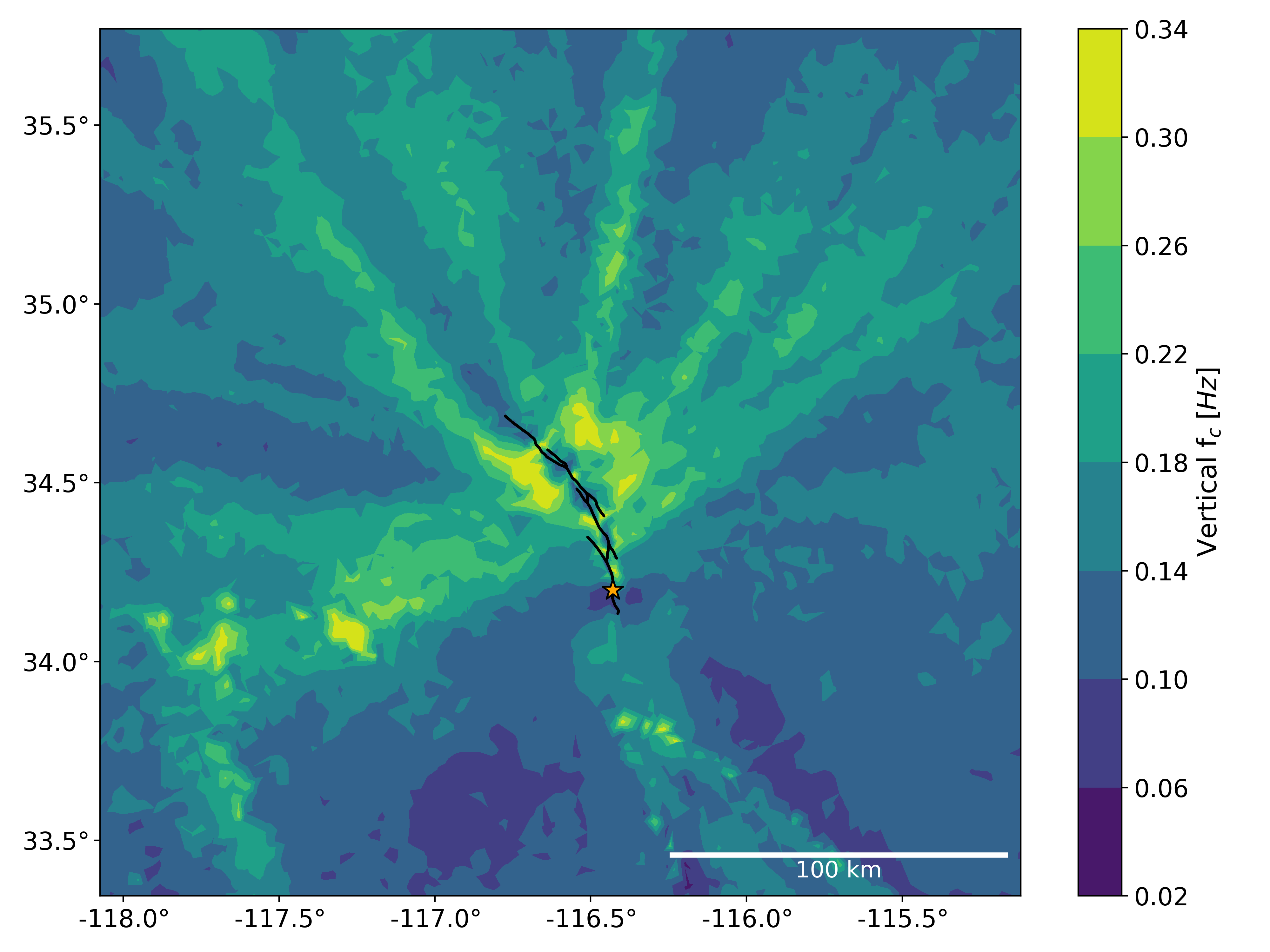}
\caption{Map view of the equivalent near-field corner frequency ($f_c$) distribution of the vertical components of synthetic seismograms simulated at $\sim$10,000 virtual seismic stations (1\% of the total virtual seismic stations). The seismograms are generated from the complex 3D dynamic rupture model of the 1992 Landers earthquake (Figure \ref{fig:landers_overview_plot}). We clip the color map at sedimentary basins and close to the fault, where static displacement and an inaccurate component separation due to finite-fault effects distort the corner frequency determination. We omit these regions in our interpretation. Black lines indicate the fault traces and the star marks the epicenter.}
\label{fig:Landers_fc_map_low_res}
\end{figure}

\end{document}